\documentclass[fleqn,usenatbib]{mnras}

\usepackage{newtxtext,newtxmath}

\usepackage[T1]{fontenc}
\usepackage{ae,aecompl}

\usepackage{graphicx}	
\usepackage{amsmath}	
\usepackage{siunitx}	
\usepackage{booktabs} 
\usepackage[table]{xcolor} 

\newcommand{\code}[1]{\textsc{\MakeLowercase{#1}}} 
\newcommand{\solmass}{\ensuremath{\mathrm{M}_{\sun}}}	
\newcommand*{\NGC}{\ensuremath{N_\mathrm{GC}}} 
\newcommand*{\MGC}{\ensuremath{M_\mathrm{GC}}} 
\newcommand*{\Mmax}{\ensuremath{M_\mathrm{max}}} %
\newcommand*{\Mmin}{\ensuremath{M_\mathrm{min}}} %
\newcommand*{\Mgas}{\ensuremath{M_\mathrm{gas}}} 
\newcommand*{\Mvir}{\ensuremath{M_\mathrm{vir}}} 
\newcommand*{\Mseed}{\ensuremath{M_\mathrm{seedGC}}} 
\newcommand*{\MDMGC}{\ensuremath{M_\mathrm{DM,GC}}} 
\newcommand*{\Msmooth}{\ensuremath{M_\mathrm{smooth}}} 
\newcommand*{\Amin}{\ensuremath{A_\mathrm{min}}} 
\newcommand*{\etaGC}{\ensuremath{\eta_\mathrm{GC}}} 
\newcommand*{\SFR}{\ensuremath{\mathrm{SFR}}} 
\DeclareSIUnit\parsec{pc} 
\DeclareSIUnit\dex{dex} 
\DeclareSIUnit\year{yr} 
\definecolor{hc}{gray}{0.75} 

\defcitealias{burkert20}{B\&F20}
\defcitealias{harris17}{H+17}

\title[Empirical Globular Cluster Numbers]{Globular cluster numbers in dark matter haloes in a dual formation scenario: an empirical model within \code{EMERGE}}

\author[Valenzuela et al.]{
Lucas M. Valenzuela,$^{1}$\thanks{E-mail: lval@usm.lmu.de}
Benjamin P. Moster,$^{1,2}$
Rhea-Silvia Remus,$^{1}$
\newauthor{
Joseph A. O'Leary,$^{1}$
and
Andreas Burkert$^{1,3,4}$
}
\\
$^{1}$Universit\"{a}ts-Sternwarte, Ludwig-Maximilians-Universit\"{a}t M\"{u}nchen, Scheinerstraße 1, D-81679 M\"{u}nchen, Germany\\
$^{2}$Max-Planck-Institut f\"{u}r Astrophysik, Karl-Schwarzschild-Straße 1, D-85748 Garching, Germany\\
$^{3}$Max-Planck-Institut f\"{u}r extraterrestrische Physik, Giessenbachstraße 1, D-85748 Garching, Germany\\
$^{4}$Excellence Cluster ORIGINS, Boltzmannstraße 2, D-85748 Garching, Germany\\
}

\date{Accepted XXX. Received YYY; in original form ZZZ}

\pubyear{2021}

\begin{document}
\label{firstpage}
\pagerange{\pageref{firstpage}--\pageref{lastpage}}
\maketitle

\begin{abstract}

We present an empirical model for the number of globular clusters (GCs) in galaxies based on recent data showing a tight relationship between dark matter halo virial masses and GC numbers.
While a simple base model forming GCs in low-mass haloes reproduces this relation, we show that a second formation pathway for GCs is needed to account for observed younger GC populations.
We confirm previous works that reported the observed linear correlation as being a consequence of hierarchical merging and its insensitivity to the exact GC formation processes at higher virial masses, even for a dual formation scenario.
We find that the scatter of the linear relation is strongly correlated with the relative amount of smooth accretion: the more dark matter is smoothly accreted, the fewer GCs a halo has compared to other haloes of the same mass.
This scatter is smaller than that introduced by halo mass measurements, indicating that the number of GCs in a galaxy is a good tracer for its dark matter mass.
Smooth accretion is also the reason for a lower average dark matter mass per GC in low-mass haloes.
Finally, we successfully reproduce the observed general trend of GCs being old and the tendency of more massive haloes hosting older GC systems. Including the second GC formation mechanism through gas-rich mergers leads to a more realistic variety of GC age distributions and also introduces an age inversion in the halo virial mass range $\log\Mvir/\si{\solmass} =$ \num{11}--\num{13}.

\end{abstract}




\section{Introduction}
\label{sec:introduction}

Globular clusters (GCs) are among the oldest objects observed in our Universe and can be found in all galaxies but the smallest dwarfs. It has generally been found that the more massive a system is, the larger the GC population in it is \citep[e.g.][]{blakeslee97}. In fact, as reported by \citet{harris17}, there exists a tight correlation between the total dark matter halo mass of a galaxy and its number of GCs. For the most massive galaxies, GC systems of more than \num{30000}~GCs have been reported \citep{harris:2016}.
On the other hand, our own galaxy, the Milky Way (MW), has a rather poor system of GCs with only about \num{150} of them \citep{harris:1996}.
Recently, \citet{forbes18} followed this trend between dark matter halo mass and GC system richness down to dwarf galaxies with total masses around \SI{e9}{\solmass}, including systems in which only one GC was found as expected from this relation.
Even for ultra-diffuse galaxies (UDGs) in the COMA galaxy cluster, GC systems of similar richness have been reported \citep{vandokkum:2017,amorisco:2018,forbes:2020}, albeit with a tendency for the UDG GC systems to be richer than those of normal dwarf galaxies of similar stellar masses \citep{vandokkum:2017,forbes:2020}. Interestingly, while estimates of the dark matter masses in UDGs are extremely difficult, the results from dynamical modelling point towards UDGs having a similar halo mass to GC richness as normal galaxies of the same dark matter mass \citep{vandokkum:2016,vandokkum:2017}. However, there is still ongoing debate about this topic, as recently \citet{saifollahi21} concluded that the number of GCs in some UDGs might actually be strongly overestimated.

While GCs have been studied in detail, especially in our own MW, their origin and formation is still a matter of debate. Measurements of the ages of GCs are extremely difficult, in particular for GCs outside the MW, but they all indicate GCs generally being old
\citep[e.g.][]{salaris:1997,cohen:1998,brodie:2005,strader:2005,brodie:2006,sharina:2006,chiessantos:2012,vandenberg:2013,leaman:2013}, at \SI{10}{\giga\year} and older. However, a few galaxies with younger populations of GCs are also reported: for example, our neighbouring galaxy, the small Magellanic Cloud, seems to have a rather young GC system with the oldest GC being only \SI{8}{\giga\year} and the youngest less than \SI{5}{\giga\year} old (\citealp{parisi:2014}, see also \citealp{sharina:2006} for young GCs in other local group dwarf galaxies).
Especially young GCs have also been reported for NGC\,1316 \citep{goodfrooij:2001,sesto:2018}.
More recently, \citet{usher:2019} measured the age distributions for three galaxies outside the local group in comparison with the MW, and found a tendency for more massive galaxies to host older GC systems, while less massive galaxies feature populations of younger globular clusters (see also \citealp{beasley:2008} for the GC system of NGC\,5128).
Similarly, \citet{chiessantos:2011} found that, on average, S0 galaxies have younger GC systems than elliptical galaxies.

These age differences are speculated to be related to the observation that GCs usually come in two different categories of colours and metallicities: metal poor, blue GCs, and metal-rich, red GCs \citep[e.g.][]{usher:2012}. Most massive galaxies have GCs of both kinds, with the metal-rich red GCs also occurring in the inner parts of galaxies and tracing the inner light distribution \citep[e.g.][]{peng:2004,schuberth:2010,pota:2013}, while the metal-poor blue GCs are usually distributed at larger radii and can have very different density slopes compared to the observed stellar light \citep[e.g.][]{forbes:1997,schuberth:2010}.
This gave rise to the idea that blue GCs are usually formed in small, metal-poor galaxies at very early times and successively assembled onto the larger galaxies \citep[e.g.][]{cote:1998,schuberth:2010,chiessantos:2011,forbes:2018}, while red, metal-rich GCs are formed later in larger galaxies when the gas has already been enriched with metals, for example during a starburst of gas-rich major mergers \citep[e.g.][]{goodfrooij:2001,schuberth:2010,harris:2017b}.
This is also supported by several observations that found that the red GCs are younger while the blue GCs are older \citep{peng:2004,beasley:2008}, even though other studies see no such strong age differences between the blue and red GC systems \citep[e.g.][]{strader:2005}.

While this idea of a dual formation scenario for GCs in galaxies has been tested extensively in observations and is still a matter of debate, GCs have proven to be rather difficult to capture from the modelling and simulation sides:
simulations of GC systems in cosmological galaxy formation simulations have recently started to produce first results on the formation and assembly of GCs in and onto galaxies \citep[e.g.][]{pfeffer:2018,kim:2018,kruijssen:2019,lahen:2019}. However, while those first simulations successfully formed GCs with metallicities and ages comparable to observed GCs, those studies usually focused on isolated galaxies and their formation pathways and until now could not account for the multiple pathways of galaxy formation responsible for the whole range of galaxies from dwarfs to brightest cluster galaxies.

Another approach has been to implement the formation of GCs in semi-analytical or empirical models of galaxy formation. Here, different approaches can be followed:
\citet{beasley:2002} used a semi-analytic model assuming that GCs are formed either in proto-galaxies at redshifts $z>5$ or through gas-rich merging of these proto-structures and thus find two populations of metal-rich and metal-poor GCs with ages similar to the observations.
\citet{choksi:2018} assumed that GCs are usually formed in massive merger events and thus used an empirical model to populate their galaxies (based on the Millennium simulation runs) with GCs in such events. They found a remarkable agreement between their results and observations in terms of ages, metallicities, and stellar masses of the GC systems with the host galaxies' halo masses.
\citet{kruijssen:2015} modelled the GC distribution in galaxies assuming that GCs are formed at redshifts $z>2$ inside high-pressure disks and redistribute later on inside the galaxies through hierarchical merging.
The idea of hierarchical merging to establish the present-day observed relation between the numbers of GCs and the total mass of a galaxy has also been studied by \citet{boylan17} in a phenomenological model, finding that the observed relations can be well reproduced by forming GCs by $z=6$ in dark matter haloes with masses above \SI{1.07e9}{\solmass}. In a subsequent study, \citet{elbadry19} showed that this GC-to-halo mass relation is a natural consequence of the central limit theorem in the case of hierarchical merging, making the richness of a galaxy's GC system a good tracer for its total halo mass.
Both \citet{boylan17} and \citet{elbadry19} make use of halo merger trees generated by the extended \citet{press74} model with the algorithm by \citet{parkinson08}, which reproduces merger trees from N-body simulations well \citep{jiang14}.
The connection to the central limit theorem was further supported by a simple hierarchical merging model by \citet{burkert20}, who showed that with an average seeding halo virial mass per GC of $M_\mathrm{seed} = \SI{5e8}{\solmass}$ and hierarchical merging, the observed linear relation between GC richness and the dark matter halo virial mass of a galaxy can be reproduced.

In this work we present an empirical model for the number of GCs in dark matter haloes, implemented within \code{EMERGE} and building on the investigations and models by \citet{boylan17}, \citet{choksi:2018}, and \citet{burkert20}, to further analyse the hierarchical nature of GC system assembly. In Section~\ref{sec:method}, we introduce the simulations, the components of the base and extended model, and the observational data used in our analysis. Section~\ref{sec:seedingmass} describes how the model was fit to the observations, matching the general relation between halo virial masses and globular cluster numbers and its scatter. In Section~\ref{sec:redshift_evolution}, we present a variety of features that can be explored in the fitted model and their dependence on redshift. Addressing the drawbacks found in the first sections for the base model, in Section~\ref{sec:dualformation} we present how an extended model with a second formation pathway for GCs through gas-rich merger events changes and improves on the previously shown results. Finally, we discuss our findings in Section~\ref{sec:discussion} and summarise the key results in Section~\ref{sec:conclusion}.

\section{Method}
\label{sec:method}

\subsection{\texorpdfstring{\code{EMERGE}}{EMERGE} \& Backbone Simulations}
\label{sec:simulation}

For the globular cluster formation model, we extracted halo merger trees from a cosmological dark matter-only N-body simulation with a side length of \SI{30}{\mega\parsec}. This simulation adopted cosmological parameters consistent with \citet{planck16}: $\Omega_m=0.308$, $\Omega_\Lambda=0.692$, $\Omega_b=0.0484$, $H_0=\SI{67.81}{\kilo\meter\per\second\per\mega\parsec}$, $n_s=0.9677$, and $\sigma_8=0.8149$.
To generate the initial conditions, \code{MUSIC} \citep{hahn11} was used with a power spectrum computed by \code{CAMB} \citep{lewis00}. The simulation contains $512^3$ dark matter particles with particle mass $m_\mathrm{DM} = \SI{7.90e6}{\solmass}$.
Using the TreePM code \code{GADGET3} \citep{springel05} with periodic boundary conditions, the simulation was run from $z=63$ to~0, creating 132~snapshots ranging from $a = \num{.0475}$ to \num{1.0000}, equally spaced in scale factor.
The halo finder \code{ROCKSTAR} \citep{behroozi13rockstar} was used to identify the dark matter haloes in each snapshot, which were linked to construct halo merger trees with \code{Consistent Trees} \citep{behroozi13trees}. In this work, the term `main halo' is used for haloes that are not contained within any larger halo, whereas `subhalo' is used for all other haloes.

We implemented the GC formation model within the code of the empirical model \code{EMERGE} \citep{moster18}, which populates the simulated haloes with galaxies to reproduce observed statistical galaxy data. For more details on the merger histories of the galaxies found in \code{EMERGE}, see \citet{oleary:2021}. In all the GC model runs, we used the \code{EMERGE} model parameters found in their table~1.

\begin{figure}
	\centering
	\includegraphics{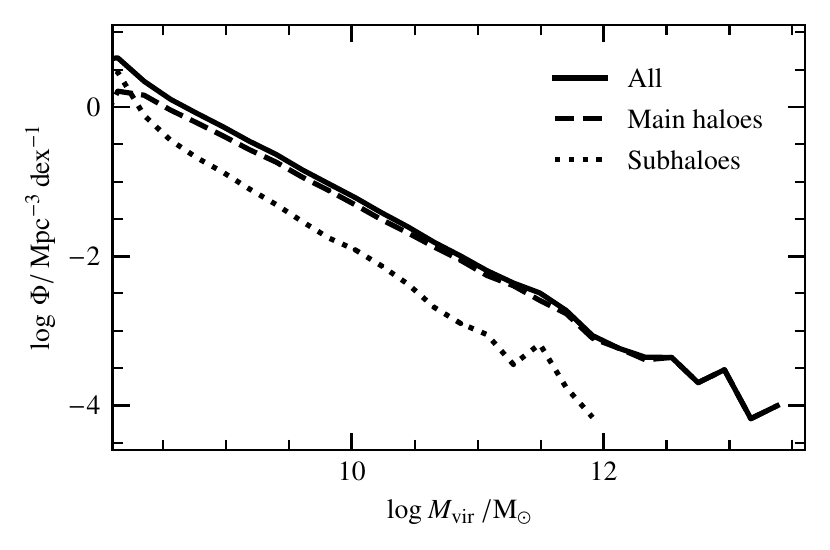}
	\caption{
		Halo mass functions of the simulation at $z = 0$ for main haloes, subhaloes, and both types together.
	}
	\label{fig:hmf}
\end{figure}

The halo mass function at $z = 0$ (Fig.~\ref{fig:hmf}) that resulted from the simulation shows that the chosen resolution allows us to make reasonable statistics for halo masses above ${\sim}\SI{e8}{\solmass}$ (see also \citealp{knebe11} for a comparison and discussion of halo finders, including their resolution limits). We can also see that the sample includes halo masses of up to almost \SI{e13.5}{\solmass}. This upper limit will suffice for comparison with the samples used by \citet{burkert20} and \citet{harris17}, which contain masses up to \SIlist{e14;e14.5}{\solmass}, respectively (excluding the galaxy clusters).

\subsection{Globular Cluster Formation Model}
\label{sec:gc_model}

Motivated by the fact that GCs come in two different flavours, i.e. red and blue GCs, we considered two GC formation models: one that we refer to as the base model, which forms GCs in small haloes, and a second one that forms GCs in-situ during gas-rich mergers.

\subsubsection{Base Model}
\label{sec:base_model}

For the base model, we used a simple phenomenological model for creating GCs in low-mass haloes, for example through gravitational collapse of a single gas cloud \citep{peebles68}. Assuming that the observed correlation between the number of GCs in a halo, \NGC{}, and the virial mass of a halo, \Mvir{}, follows from hierarchical merging, the goal is to find the halo virial mass at which GCs have to be seeded, \Mseed{}, to reproduce observations at $z = 0$. We made some simplifying assumptions for the seeding process to keep the number of free parameters at a minimum:
(1) Each halo has a one-time opportunity to form GCs as soon as it reaches \Mseed{} for the first time. (2) GCs are seeded according to a predetermined distribution $p(x)$ with expectation value~\num{1}.
Additionally, two further assumptions were made with respect to already existing GCs: (3) A halo that becomes a subhalo immediately turns over all its GCs to the respective main halo. (4) A GC cannot disappear. The latter assumptions are introduced to avoid additional model parameters, but these aspects provide good starting points for future refinements and expansions of the model.

From assumption~1 it follows that merging events do not have a direct influence on the formation of GCs in this model and that a distribution of different GC ages is obtained.
For assumption~2, we tested three different seeding functions: the first seeding distribution implemented, $p_1(x)$, simply seeds exactly one GC in every seeding process. The second distribution, $p_2(x)$, seeds zero, one, or two GCs with equal probability. Finally, the third distribution we tested is a so-called `geometric distribution', $p_3(x) = 1/2^{x+1}$, which seeds GCs according to an exponential law.
Assumption~3 leads to the possibility of backsplash haloes having very few or no GCs at all. This does not affect our conclusions because all the fits were performed on the simulated main haloes with at least one GC. Even when considering all main haloes, the difference at high masses is negligible (e.g.\ Fig.~\ref{fig:mdmgc}).
Assumption~4 means that GC disruption is neglected in our model. Note that for example for the relation between GC mass and halo virial mass, \citet{bastian20} come to the conclusion that GC disruption plays a dominant role, though for example \citet{burkert20} point out that GC disruption does not affect the linear correlation between \Mvir{} and \NGC{}.

\subsubsection{Model Extension}
\label{sec:model_extension}

With the aim to address issues found in the results of the base model and motivated by the suggestion that the observed blue and red GC populations may originate from different formation scenarios \citep[e.g.][]{cote:1998,schuberth:2010}, we added an extension to the base model (discussed in Section~\ref{sec:dualformation}): a second formation mechanism triggered by gas-rich merger events \citep[e.g.][]{ashman92}. To keep the number of additional parameters to a minimum and taking into account the framework that \code{EMERGE} provides, we decided to follow the approach used by \citet{choksi:2018} for the GC formation model. GC formation is triggered on the condition that the logarithmic halo mass accretion rate, $A_m$, surpasses a certain threshold, \Amin{}, which is generally the case for major mergers:
\begin{equation}
    \label{eq:p3}
    A_m = \frac{M_\mathrm{vir} - M_\mathrm{vir,prog}}{M_\mathrm{vir} \Delta t} > \Amin,
\end{equation}
where $M_\mathrm{vir}$ and $M_\mathrm{vir,prog}$ are the virial masses of the halo considered for GC formation and its progenitor, respectively, $\Delta t$ is the timestep between their two snapshots, and the threshold \Amin{} is one of the two free parameters of the model. Note that this corresponds to the parameter $p_3$ of \citet{choksi:2018}.

When the formation criterion is met, the mass of the present cold gas is computed according to $M_\mathrm{gas} = \SFR \times t_\mathrm{dep}$, where \SFR{} is the star formation rate computed by \code{EMERGE} \citep[according to the best fit to observations,][]{moster18,oleary:2021} and $t_\mathrm{dep}$ is the gas depletion time, which we calculate from the redshift according to \citet{tacconi20}:
\begin{equation}
    \label{eq:tdep}
    t_\mathrm{dep} = \varepsilon_H^{-1} \times t_H \times \big(\num{.7} + \num{.3} \times (1+z)^3 \big)^{-1/2},
\end{equation}
with $\varepsilon_H^{-1} = \num{.1}$ and $t_H = \SI{13.98}{\giga\year}$. The total mass of GCs formed in the formation event, \MGC{}, is then determined from the cold gas mass, \Mgas{}:
\begin{equation}
    \label{eq:p2}
    \MGC = \num{1.8e-4} \etaGC \Mgas,
\end{equation}
where \etaGC{} is the second and last free parameter of the model, corresponding to $p_2$ of \citet{choksi:2018}. The same conversion was previously also applied by \citet{li14}, based on predictions by \citet{kravtsov05}. We adapted the same cluster initial mass function and normalisation as \citet{choksi:2018},
\begin{equation}
    \label{eq:cimf}
    \frac{dN}{dM} = \Mmax M^{-2},
\end{equation}
with \Mmax{} being the most massive GC formed. \Mmax{} depends on \MGC{} according to
\begin{equation}
    \label{eq:MGC}
    \MGC = \int_{\Mmin}^{\Mmax} dM M \frac{dN}{dM} = \Mmax \ln\frac{\Mmax}{\Mmin},
\end{equation}
where $\Mmin = \SI{e5}{\solmass}$ is the minimum GC mass that is expected to survive for a few \si{\giga\year} \citep{li14,choksi:2018}. This can be solved for $\Mmax/\Mmin$ with the Lambert~$W$ function:
\begin{equation}
    \label{eq:MmaxMmin}
    \frac{\Mmax}{\Mmin} = \exp\bigg(W\Big(\frac{\MGC}{\Mmin}\Big)\bigg),
\end{equation}
from which the average number of formed GCs can finally be obtained:
\begin{equation}
    \label{eq:NGC}
    \langle\NGC\rangle = \int_{\Mmin}^{\Mmax} dM \frac{dN}{dM} = \frac{\Mmax}{\Mmin} - 1.
\end{equation}
Like for the approach of \citet{elbadry19}, if an average of for example \num{5.2}~GCs would be formed, the halo forms 5 GCs with a probability of 80~per cent and 6 GCs otherwise. No GCs are formed if $\MGC < \Mmin$.

\subsection{GC Studies \& Obtaining Mock Observations}
\label{sec:mockobservations}

Having specified our model, the free parameter \Mseed{}, and the available seeding distributions, we populated the haloes from the merger trees with GCs using the base model. To obtain the best value for \Mseed{} and its uncertainty, the computed numbers of GCs per halo were compared to the observed samples used by \citet{burkert20} and \citet{harris17} (hereafter \citetalias{burkert20} and \citetalias{harris17}).
Both are compilations of literature work, where the former sample focuses on high-quality studies, i.e., the corrections used to determine \NGC{} are required to be small (see \citealp{harris15} for a brief overview of the typical approach to getting \NGC{}) and includes a variety of low-mass galaxies. The latter study includes a larger sample size, of which the large majority have masses above \SI{e11}{\solmass}. Because there are overlaps between the two studies and to prevent systematic differences in their approaches of obtaining \NGC{} and \Mvir{} from distorting our results, we decided to fit our model to the two observational studies separately.
For the comparison with the former sample, we used the mean observed values of $\log\NGC$ from their equation~1,
\begin{equation}
    \label{eq:NGCMvirlogburkert}
    \langle \log\NGC \rangle = (\num{-9.58+-1.58}) + (\num{.99+-.13}) \times \log\Mvir/\solmass,
\end{equation}
and the scatters, $\sigma_\mathrm{obs}$, in the mass bins used in fig.~2 of \citetalias{burkert20}. For the sample from \citetalias{harris17}, we determined the mean linear correlation,
\begin{equation}
    \label{eq:NGCMvirlogharris}
    \langle \log\NGC \rangle = (\num{-8.86+-0.89}) + (\num{.91+-.07}) \times \log\Mvir/\solmass,
\end{equation}
and the scatters from it in the same bins.

To get mock observations of the simulated virial masses, we added normally distributed error values to their logarithms (see Section~\ref{sec:directoutput} about how the variance of the error distribution was determined). We then binned the haloes in mock virial mass and calculated the mean of $\log\NGC$ and its scatter, $\sigma_\mathrm{sim}$, in each bin. The scatter was calculated as the $1\sigma$-deviation of $\log\NGC - \langle \log\NGC \rangle$ within a bin. Using a spline interpolation, we could then compute the values for each observed data point.
To quantify the likelihood of the chosen parameters, we used the $\chi^2$ goodness-of-fit test.

\subsection{Fitting Model Parameters}
\label{sec:fittingparameters}

To determine the parameter \Mseed{}, we required the model to reproduce the observed relation between \Mvir{} and \NGC{}. To this end, we used a Markov chain Monte Carlo (MCMC) method, taking into account the best mock observation error distribution (see Sections~\ref{sec:directoutput} and~\ref{sec:mockoutput}).
The sampler is implemented according to one of the methods described by \citet{goodman10}, an affine invariant ensemble sampler for MCMC (see \citealp{moster18} for more details on the implementation). \Mseed{} was the only free parameter, with all other \code{EMERGE} model parameters held fixed.

\section{Globular Cluster Seeding Mass}
\label{sec:seedingmass}

\subsection{Direct Simulation Results}
\label{sec:directoutput}

\begin{figure}
	\centering
	\includegraphics{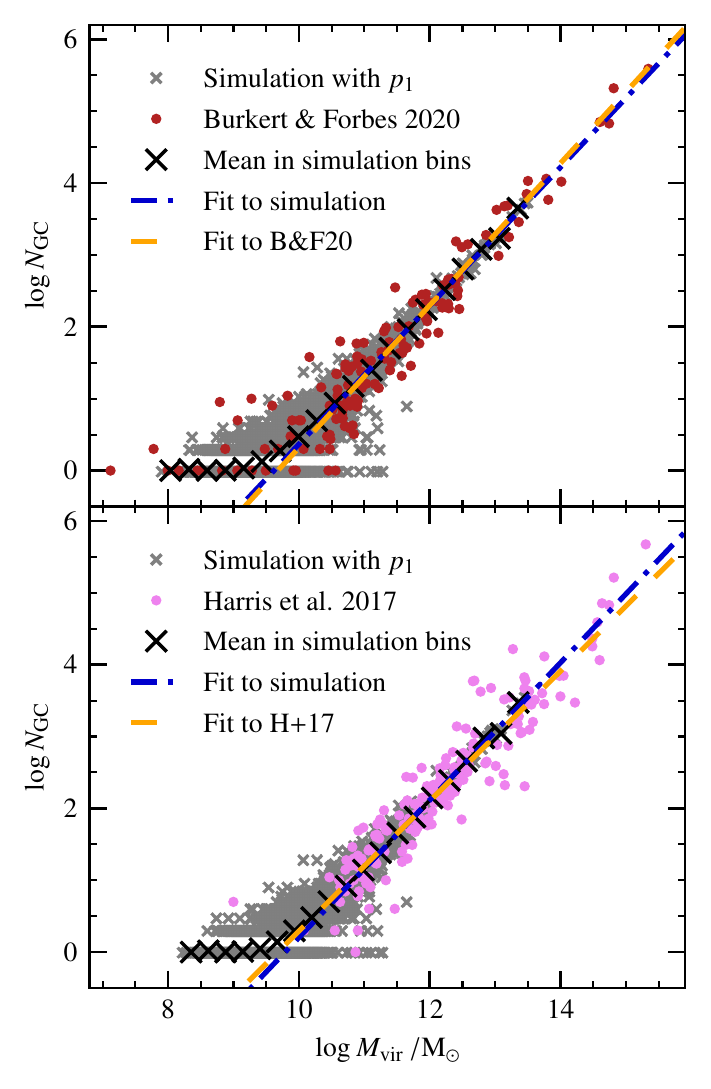}
	\caption{
		Correlation between dark matter halo virial masses, \Mvir{}, and numbers of globular clusters, \NGC{}, of the simulated haloes compared to the observed galaxies from \citetalias{burkert20} (top) and \citetalias{harris17} (bottom). GCs were seeded at a virial mass of \SIlist{e8.67;e8.86}{\solmass}, respectively. The distribution used for the number of GCs to seed was $p_1$, where exactly one GC is seeded in a seeding event. The fits to the simulations were computed by the least-square method for the bin masses above \SI{e10.5}{\solmass}. The fits to the observations correspond to Equations~\ref{eq:NGCMvirlogburkert} and~\ref{eq:NGCMvirlogharris}, respectively.
	}
	\label{fig:MvirNGC1}
\end{figure}

As a first step, we manually fitted the obtained GC numbers with the seeding distribution $p_1$ to the observed ones by adjusting \Mseed{} without using the MCMC method. This way, we were able to obtain linear correlations that are very similar to the observed ones (Fig.~\ref{fig:MvirNGC1}). Note that the virial masses here are not the masses that result from mock observations, but the original ones from the merger trees. The best results were obtained with $\Mseed = \SI{4.7e8}{\solmass}$ and $\Mseed = \SI{7.2e8}{\solmass}$ for the samples of \citetalias{burkert20} and \citetalias{harris17}, respectively. After putting the haloes with at least one GC at $z = 0$ into 20~mass bins, we calculated the mean of $\log\NGC$ for each of them. The mean value becomes near linear when plotted against $\log\Mvir{}$ around \SI{e10.5}{\solmass}, so we computed the least-square fitting line for the haloes above that mass, which resulted in the following respective relations:
\begin{align}
\langle \log\NGC \rangle &= \num{-9.27} + \num{0.96} \times \log\Mvir/\solmass, \label{eq:NGCMvirlog1} \\
\langle \log\NGC \rangle &= \num{-9.36} + \num{0.96} \times \log\Mvir/\solmass.  \label{eq:NGCMvirlog1_harris}
\end{align}
Both y-intercepts and slopes are well within the error ranges of the fits to the observed samples.

\begin{figure}
\centering
\includegraphics{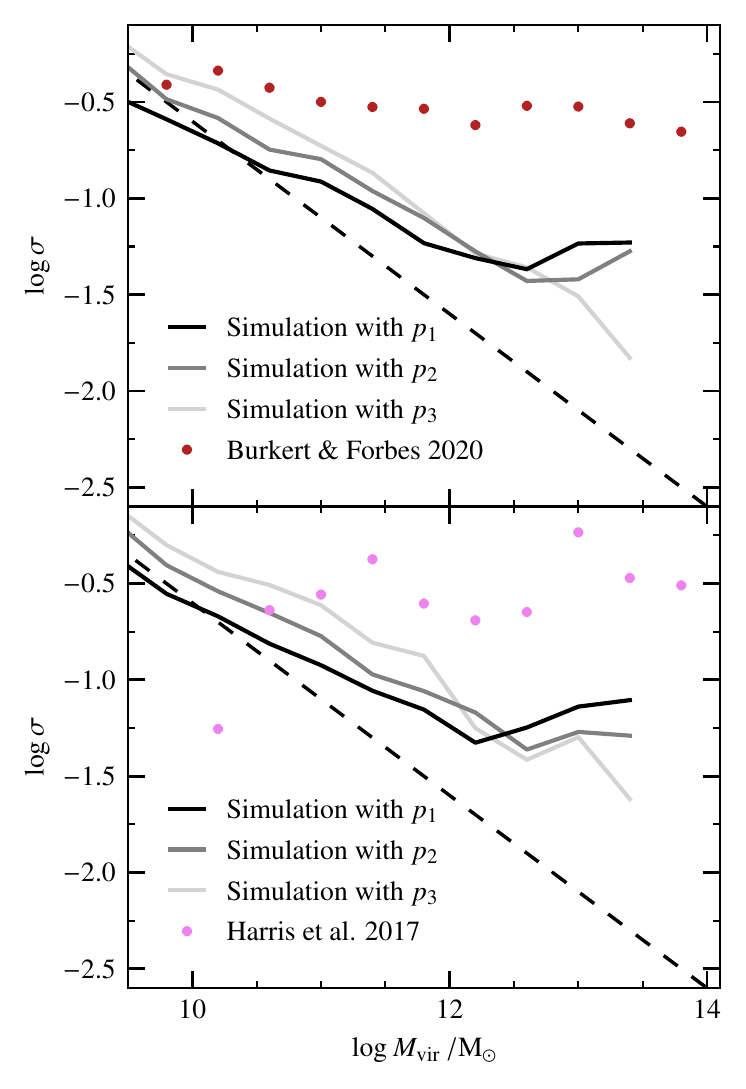}
\caption{
	Scatter of \NGC{} around the mean correlation compared between the simulated haloes and observations from \citetalias{burkert20} and \citetalias{harris17}. GCs were seeded at a virial mass of \SIlist{e8.67;e8.86}{\solmass} (top and bottom, respectively). The scatter of the simulated values decreases at a smaller rate than $\sigma \propto 1 / \sqrt{\Mvir}$ (dashed line), which corresponds to the prediction of the central limit theorem \citepalias{burkert20}. Neither the different GC seeding distributions nor the two values of \Mseed{} have a significant effect on the slope. In contrast, the observed values from the sample of \citetalias{burkert20} have a near constant scatter around the mean correlation. The sample of \citetalias{harris17} follows a very similar correlation, although the scatter varies more around it.
}
\label{fig:Mvirsigma1}
\end{figure}

However, one can clearly see that the scatter in $\log\NGC$ for a given mass is much larger for the observed values than for the simulated ones, in particular at high virial masses. While it decreases strongly with increasing mass for the simulated haloes, it remains roughly constant for the observed values (Fig.~\ref{fig:Mvirsigma1}). This was first noticed by \citetalias{burkert20} when they compared their observational sample to the results of their simple Monte Carlo halo merging simulation.

\begin{figure}
\centering
\includegraphics{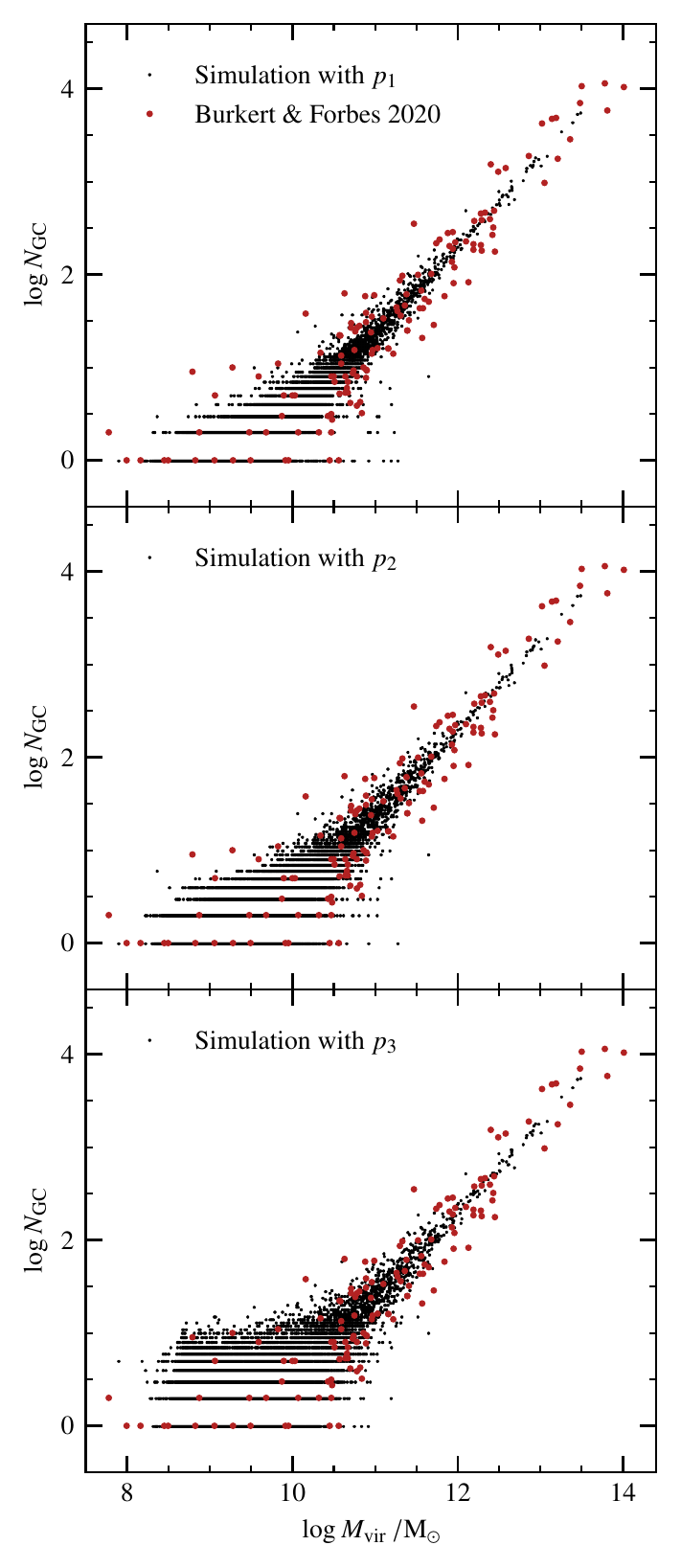}
\caption{
	Correlation between dark matter halo virial masses, \Mvir{}, and numbers of globular clusters, \NGC{}, of the simulated haloes compared to the observational sample used by \citetalias{burkert20}. GCs were seeded at a virial mass of \SI{e8.67}{\solmass}. The distributions used for the numbers of GCs to seed were $p_{1,2,3}$, from top to bottom, respectively. Note that the scatter of \NGC{} is larger in $p_2$ and $p_3$ for low masses, but almost identical to the one in $p_1$ for masses above ${\sim}\SI{e11}{\solmass}$. The same result is also found for the sample from \citetalias{harris17}.
}
\label{fig:MvirNGCall}
\end{figure}

There are two straightforward ways of changing the scatter in our model: manipulating the scatter of \NGC{} and manipulating the scatter of \Mvir{}. The former can be addressed by using one of the other seeding distributions, $p_2$ or $p_3$, since they have a larger variance than $p_1$. However, when comparing the scatter of the three resulting distributions of GCs in the haloes, we found only a slight increase of the scatter with $p_2$ and $p_3$, but no significant flattening (Fig.~\ref{fig:Mvirsigma1}).
\citetalias{burkert20} argued that through hierarchical merging the central limit theorem would predict the scatter of $\log\NGC$ to follow the relation $\sigma \propto 1 / \sqrt{\Mvir}$ (dashed line in Fig.~\ref{fig:Mvirsigma1}), which their simulation closely resembles.
Interestingly, in our simulated cases the scatter decreases at a smaller rate than the prediction of the central limit theorem.
We suggest that this may be due to the variance in smoothly accreted dark matter for haloes (discussed in Section~\ref{sec:smoothaccretion}), which could also be the reason for the flattening of the scatter curve at large virial masses by creating a lower limit for the inherent scatter of \NGC{}.
In Fig.~\ref{fig:MvirNGCall} one can see that the seeding distribution only visibly affects the scatter of \NGC{} at virial masses below ${\sim}\SI{e11}{\solmass}$. The distribution above this mass remains the same, regardless of how exactly the GCs were seeded. This agrees very well with the results from \citet{elbadry19}, who stated that for $\Mvir(z = 0) \gtrsim \SI{e11.5}{\solmass}$, the ratio between the halo virial mass and the total GC mass in the respective halo is constant, independent of the GC-to-halo mass relation at the time of GC formation.

\begin{figure}
	\centering
	\includegraphics{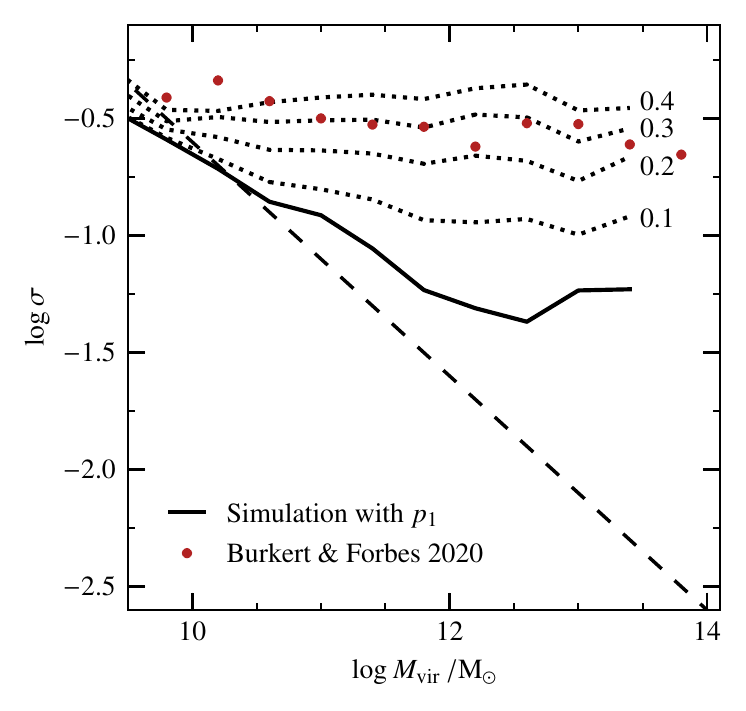}
	\caption{
		Scatter of \NGC{} around the mean correlation compared between the simulated and observed haloes from \citetalias{burkert20}. GCs were seeded at a virial mass of \SI{e8.67}{\solmass} and the seeding distribution $p_1$ was used. The dashed line corresponds to the prediction of the central limit theorem. The dotted lines show the average scatter for 1000~mock observations of the simulated virial masses, where the numbers to their right are the respective mock observation errors in \si{\dex}.
	}
	\label{fig:Mvirsigmadex}
\end{figure}

This clearly does not resolve the discrepancy between the simulated and observed scatter at high virial masses. We therefore took a look at the second possibility, following \citetalias{burkert20}: manipulating the scatter of the virial masses. We addressed this by introducing an observation error to the virial mass to obtain mock observations, as described in Section~\ref{sec:mockobservations}. The random error is drawn from a normal distribution and added to $\log\Mvir$. We tested a variety of variances for the random errors, which gave us significantly different results for the scatter (Fig.~\ref{fig:Mvirsigmadex}). We clearly get the best match for an observation error of $\delta\log\Mvir/\solmass = \SI{.3}{\dex}$, in particular for the scatter above a mass of \SI{e11}{\solmass}. This is also the range we are interested in since the linear relationship in Fig.~\ref{fig:MvirNGC1} was fitted to the haloes with virial masses above \SI{e10.5}{\solmass}. We also found the same result when comparing the model to the observational sample of \citetalias{harris17}. The value is slightly higher than the findings of \citetalias{burkert20} who inferred an error of \SI{.25}{\dex}. The logarithmic error of \SI{.3}{\dex} corresponds to an error of a factor~\num{2.0} for determining virial masses, which is a realistic value for true observations. This value is used to obtain mock observations for fitting \Mseed{} with the MCMC method.

\subsection{Mock Observations of the Simulation Results}
\label{sec:mockoutput}

Having determined the mock observation error that allows us to reproduce the observed scatter of \NGC{}, we ran the MCMC model (Section~\ref{sec:fittingparameters}) with the seeding distribution $p_1$ to get the most likely value for the parameter \Mseed{} and its error for both observational samples. We set up 100~walkers around the expected best fit and evolved each of them for 130~steps. The error of the walker distribution had converged by that time and could be extracted from the final walker samples.

\begin{figure}
	\centering
	\includegraphics{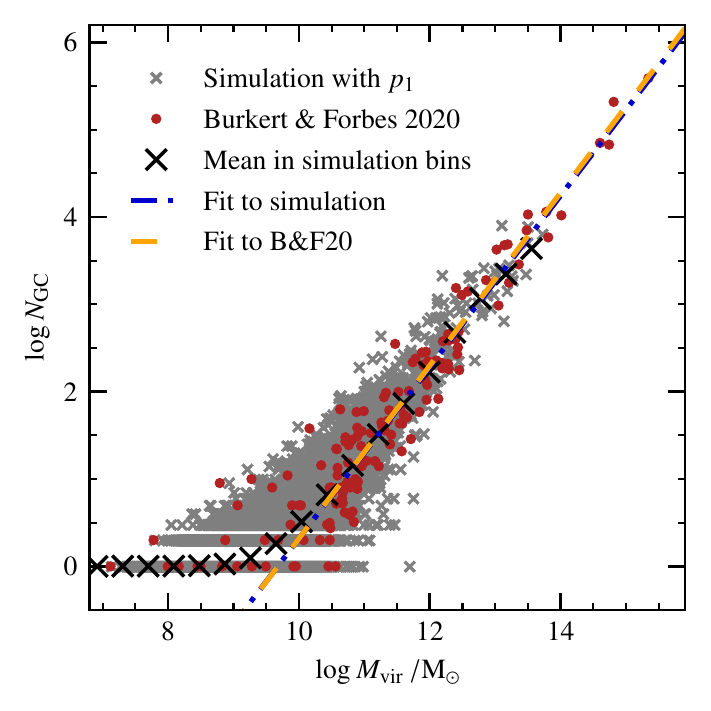}
	\caption{
		Correlation between dark matter halo virial masses \Mvir{} and numbers of globular clusters \NGC{} of the simulated haloes compared to the observational sample from \citetalias{burkert20}. GCs were seeded at a virial mass of \SI{e8.50}{\solmass} and the seeding distribution $p_1$ was used. The values of the simulated virial masses are mock observations of the originally simulated values. The fit to the simulation was computed by the least-square method for the bin masses between \SIlist{e10.5;e13}{\solmass}, averaged over 1000~mock observations of the simulated virial masses. The fit to the observations corresponds to Equation~\ref{eq:NGCMvirlogburkert}.
	}
	\label{fig:MvirNGC1mock}
\end{figure}

For \citetalias{burkert20}, we obtained a best fit for $\log\Mseed/\solmass = \num{8.50}\substack{+\num{.19} \\ -\num{.29}}$\,, with the uncertainties being the upper and lower $1\sigma$ parameter errors. For \citetalias{harris17}, the best fit was found for $\log\Mseed/\solmass = \num{8.67}\substack{+\num{.21} \\ -\num{.16}}$\,. We then ran the simulation with $\Mseed = \SI{e8.50}{\solmass}$, again adding mock observation errors to the halo virial masses, and placed the haloes with at least one GC at $z = 0$ in the mass bins. As in Section~\ref{sec:directoutput}, we calculated the mean of $\log\NGC$ in each of them and computed the least-square fitting line for the haloes with $\Mvir > \SI{e10.5}{\solmass}$, where the mean value becomes near linear (Fig.~\ref{fig:MvirNGC1mock}).
In comparison to Figs.~\ref{fig:MvirNGC1} and~\ref{fig:MvirNGCall}, the scatter of the simulated number of GCs matches the scatter of the observed galaxies much better when using mock observations, as expected from Fig.~\ref{fig:Mvirsigmadex}. Using one of the other two seeding distributions, $p_2$ or $p_3$, we can see the same effect as in Fig.~\ref{fig:MvirNGCall}: The exact seeding process is irrelevant for the distribution of the plotted simulated haloes with virial masses above \SI{e11}{\solmass}.
The same conclusions are drawn for the observational sample from \citetalias{harris17}.

Even having the mock observation errors, it is not yet possible to tell which of the seeding distributions is best for describing the observations at lower masses. We decided to use $p_2$ for the rest of this work, with the reason that it is not realistic that exactly one GC will form in every single dark matter halo with the virial mass \Mseed{}. At the same time, it is also not realistic that a theoretically unlimited amount of GCs can form in a halo of that mass, which is why we also decided against $p_3$. Since the results obtained for the fits to both observational samples are qualitatively the same, we only show the plots featuring the simulated globular cluster numbers fitted to the sample used by \citetalias{burkert20} in the following.

\section{Evolution with Redshift}
\label{sec:redshift_evolution}

\subsection{Correlation at Higher Redshifts}
\label{sec:predictions}

\begin{figure}
	\centering
	\includegraphics{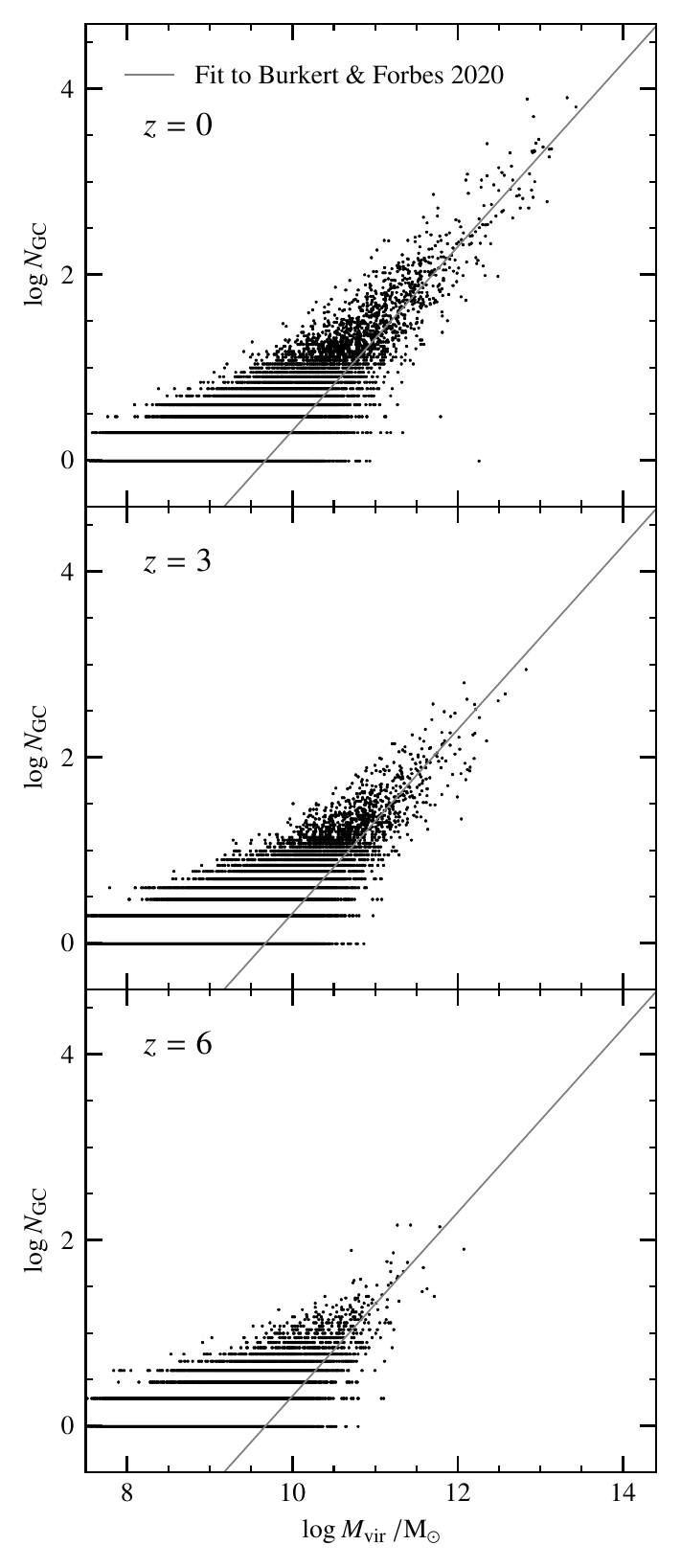}
	\caption{
		Correlation between dark matter halo virial masses, \Mvir{}, and numbers of globular clusters, \NGC{}, of the simulated haloes at redshifts $z = 0,~3, \text{ and } 6$, from top to bottom respectively, compared to the observed linear relation determined by \citetalias{burkert20} (Equation~\ref{eq:NGCMvirlogburkert}). GCs were seeded at a virial mass of \SI{e8.50}{\solmass} and the seeding distribution $p_2$ was used. The values of the virial masses are mock observations of the originally simulated values.
	}
	\label{fig:MvirNGC012mockallz}
\end{figure}

Having established a value for \Mseed{}, a mock observation error, and a seeding distribution, we determined our model's prediction of the distribution of GCs in galaxies at higher redshifts. For this, we picked three redshift values, $z = 0$,~3, and~6 (Fig.~\ref{fig:MvirNGC012mockallz}). The shape at low virial masses remains the same at any of the selected redshifts. Only the upper limit of \NGC{} and \Mvir{} is lower for higher redshifts, which is to be expected since high virial masses require many merging events, so these did not yet exist at high $z$ in the simulated box.
Even at the higher redshifts the scatter of \NGC{} around the observed linear relation matches that of the observations very well, just as for $z = 0$. Our model clearly predicts that on average, the number of GCs and its scatter will be the same in haloes of equal virial mass, independent of redshift. 

\subsection{Dark Matter Halo Mass per GC}
\label{sec:dwarfgalaxyregime}

\begin{figure}
	\centering
	\includegraphics{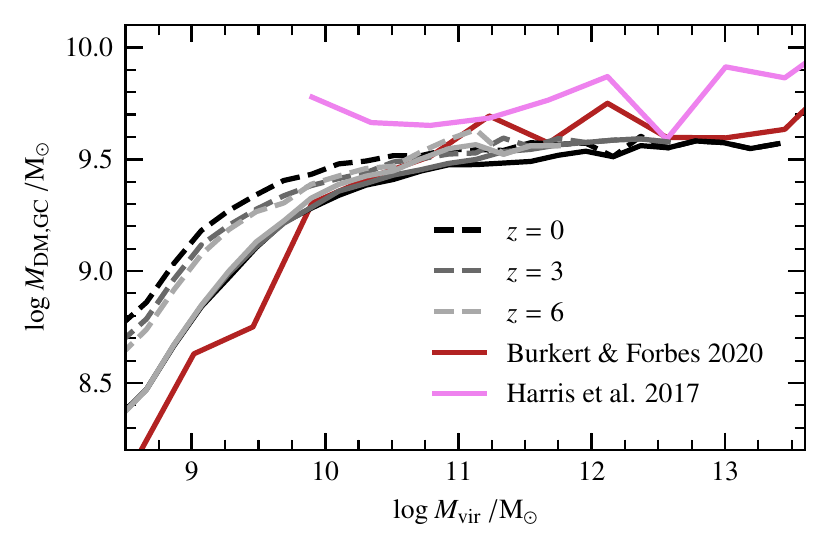}
	\caption{
		Characteristic dark matter mass in virialised haloes per GC at redshifts $z = 0$, 3, and~6, compared to the samples from \citetalias{burkert20} and \citetalias{harris17}. The dashed lines include haloes without GCs, whereas the solid lines only consist of haloes with at least one GC. GCs were seeded at a virial mass of \SI{e8.50}{\solmass} and the seeding distribution $p_2$ was used.
	}
	\label{fig:mdmgc}
\end{figure}

\citetalias{burkert20} introduced the characteristic dark matter mass in virialised haloes per GC at $z = 0$ as the arithmetic mean, $\MDMGC = \langle \Mvir/\NGC \rangle$, for haloes with at least one GC. They also remark that the linear correlation breaks down for haloes with virial masses below \MDMGC{} and predict that it extends statistically down into the dwarf galaxy regime. Since our model forms GCs well below this value, at \SI{e8.50}{\solmass}, we needed to find a quantification that addresses this statistical extension. For that, we redefined \MDMGC{} as the harmonic mean, such that haloes with no GCs in them can be included:
\begin{equation}
\label{eq:mdmgc}
\MDMGC = \left\langle \frac{\NGC}{\Mvir} \right\rangle^{-1}.
\end{equation}
Using this definition, we binned the haloes in virial mass and computed the value for \MDMGC{} in every bin at three different redshifts, $z = 0,~3, \text{ and } 6$ (Fig.~\ref{fig:mdmgc}).
When only considering haloes with at least one GC, there is no significant difference at different redshifts. Unsurprisingly, taking all haloes into account results in higher values of \MDMGC{}, particularly visible for virial masses below ${\sim}\SI{e10.5}{\solmass}$, which appears to also decrease slightly with redshift. This may be caused by a higher number of low-mass backsplash haloes with no GCs at lower redshift.
A feature seen in all cases is the decrease of \MDMGC{} towards lower virial masses below ${\sim}\SI{e10}{\solmass}$, which means that the correlation between \Mvir{} and \NGC{} is not entirely linear across the entire mass range. This effect is likely caused by `smooth accretion', which we define as any dark matter mass not accreted through mergers with haloes above \Mseed{}, therefore generally adding no GCs to the main halo (this is discussed in more detail in Section~\ref{sec:smoothaccretion})\footnote{This definition of `smooth accretion' is not physically motivated, but arises from the GC model itself to quantify the amount of dark matter accreted containing no potential GCs. It is for this reason that it breaks down at low virial masses near \Mseed{}.}. Note that there is no significant difference when using a different seeding distribution, except that when only considering haloes with at least one GC, the value of \MDMGC{} is higher at low virial masses for $p_1$ and lower for $p_3$.

The values of \MDMGC{} for haloes with at least one GC match the values of the observational sample used by \citetalias{burkert20} well, even at low virial masses, which is remarkable because the focus of the fit was the linear correlation seen at higher masses. Note that the small number of low-mass haloes in the observational sample lead to large uncertainties in that region, however. The sample from \citetalias{harris17} features an overall higher value of \MDMGC{}, which is to be expected because of the larger GC seeding mass necessary to fit the model to it. Unsurprisingly, using the simulations with $\Mseed = \SI{e8.67}{\solmass}$ shifts the simulated lines upwards to better match that sample.

\subsection{Smooth Accretion}
\label{sec:smoothaccretion}

\begin{figure}
	\centering
	\includegraphics{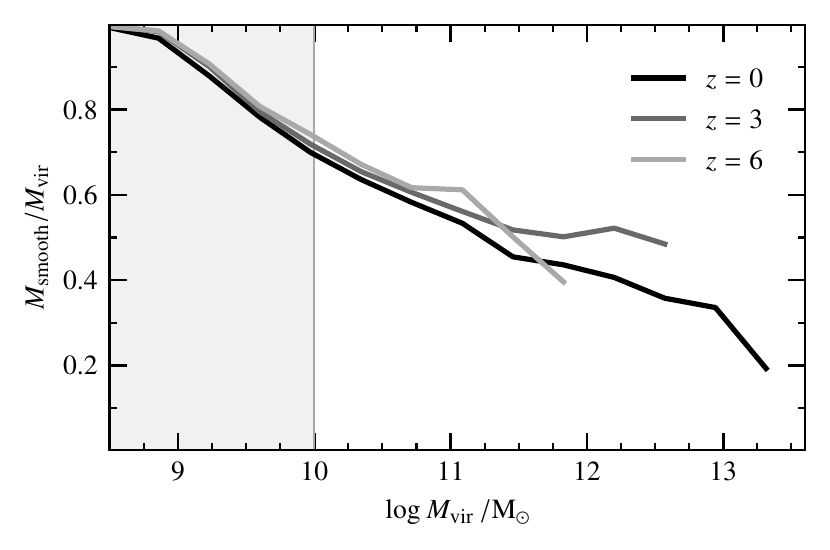}
	\caption{
	    Smoothly accreted dark matter mass fraction of the total halo virial mass of the simulated haloes at redshifts $z = 0$, 3, and~6. Because we define smooth accretion to be any mass below \Mseed{} that is accreted, this leads to resolution effects below ${\sim}\SI{e10}{\solmass}$ (grey shaded area).
	}
	\label{fig:MvirMsmooth}
\end{figure}

To take a closer look at the smoothly accreted mass in haloes, we determined the total mass accreted through mergers with haloes more massive than \Mseed{} by iterating the merger trees along the most massive progenitors. For this, we removed all substructures from the trees by immediately adding subhaloes to the respective main haloes. The remaining mass of a halo, \Msmooth{}, is considered to have been \emph{smoothly accreted}. It is therefore the mass that remains when the sum of all mergers above \Mseed{} is subtracted from a halo's mass at $z = 0$. The average fraction of the total virial mass that is smoothly accreted is shown in Fig.~\ref{fig:MvirMsmooth} as a function of virial mass. At virial masses below \SI{e10}{\solmass}, our definition of smooth accretion breaks down due to resolution effects near \Mseed{}. The overall trend shows that a larger fraction of \Mvir{} of low-mass haloes was smoothly accreted compared to that of high-mass haloes. There appears to be a small increase of the smoothly accreted mass fraction at virial masses above ${\sim}\SI{e11}{\solmass}$ at higher redshifts, which could be related to them having to grow faster in mass via larger mergers to reach the respective virial mass at earlier times.
Note that there are inherent difficulties of extracting accurate information from halo merger trees, such as consistency problems caused by fly-throughs or losing track of haloes between snapshots. Since we only iterate the trees without following the individual particles, we are not able to identify the loss of matter if the total virial mass increases from one snapshot to the next. For this reason, it is possible that the total mass accreted by mergers with haloes above \Mseed{} surpasses the total virial mass (in which case we assume \Msmooth{} to be zero), reflecting a bias towards low values of \Msmooth{}.
Still, the method suffices to find an important trend with globular cluster numbers.

\begin{figure}
	\centering
	\includegraphics{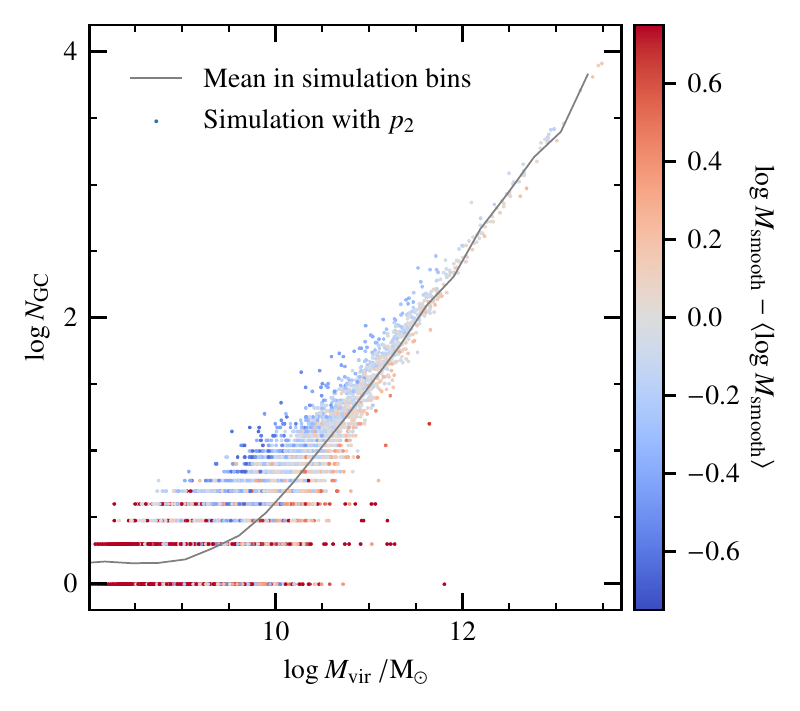}
	\caption{
	    Correlation between dark matter halo virial masses, \Mvir{}, and numbers of globular clusters, \NGC{}, of the simulated haloes, coloured by the deviation of the total smoothly accreted dark matter mass from the expected value at the respective virial mass. GCs were seeded at a virial mass of \SI{e8.50}{\solmass} and the seeding distribution $p_2$ was used.
	}
	\label{fig:MvirNGCcolored}
\end{figure}

The intrinsic scatter of \NGC{} at fixed \Mvir{} is correlated with the deviation of smoothly accreted mass from the mean at \Mvir{}, without including mock observation errors (Fig.~\ref{fig:MvirNGCcolored}). This relation can be described as follows: The more dark matter is smoothly accreted, the fewer GCs a halo has compared to other haloes of the same mass. However, for this scatter to be observable, it will first have to be possible to determine dark matter virial masses with greater accuracy, since this is still the dominant cause of scatter in observations.

Our definition of smooth accretion is comparable to the \emph{diffuse accretion} introduced by \citet{fakhouri10}, for which they either count none of the resolved haloes that are accreted or set a limit via the merger mass fraction, $\xi_\mathrm{limit} = \num{.04}$. It is for this reason that they find larger diffusely accreted mass fractions compared to this work, never dropping below 40~per cent. This is compatible with our findings because our absolute limit of \Mseed{} allows the merger mass fraction to become much smaller for high-mass main haloes, leading to fewer mergers being included in the smooth accretion. However, a direct comparison with their findings is complicated by the different quantity considered: the overdensity within the surroundings instead of \Mvir{}. Still, the general trend of a decreasing smooth accretion fraction with increasing overdensity matches the one we find with \Mvir{} very well. \citet{genel10} perform a more detailed analysis by looking at the individual dark matter particle fractions coming from mergers versus being smoothly accreted or stripped from another halo. Particles of the latter two types would roughly correspond to our smoothly accreted mass. Taking those two together, they also obtain higher values of the smoothly accreted mass fraction, above 50~per cent for high-mass haloes. In this case, their lower resolution leads to more unresolved mergers being counted towards smooth accretion.
As they state, their results of close to 100~per cent at lower masses towards \SI{e10}{\solmass} are due to resolution effects, much like we found for our analysis near the smooth accretion limit \Mseed{}. Their results do not differ across the redshift range they looked at, $z = 0\text{--}2$, which is in agreement with the small differences seen in our simulations in the range $z = 0\text{--}6$ (Fig.~\ref{fig:MvirMsmooth}). They also find a decrease of the smooth accretion fraction with \Mvir{} as we do.

\subsection{Age Distributions}
\label{sec:ages}

\begin{figure}
	\centering
	\includegraphics{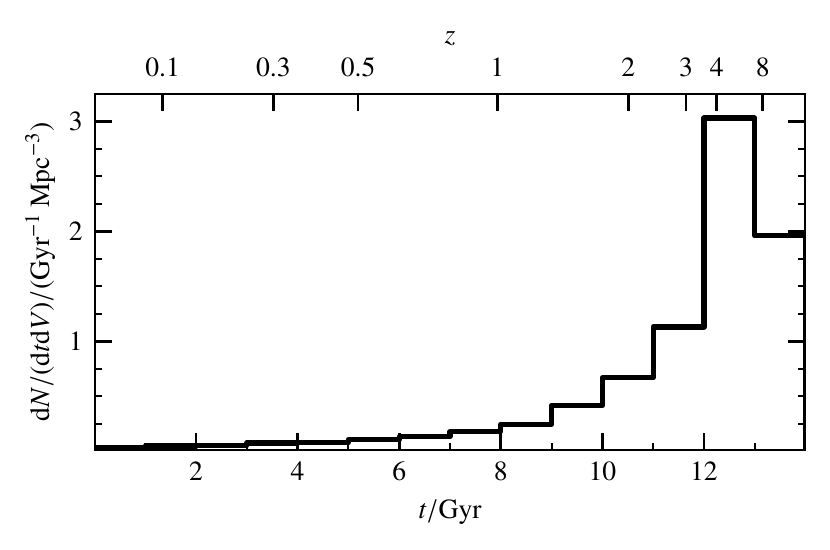}
	\caption{
	    Age distribution of GCs in the simulation, normalised to time and volume. GCs were seeded at a virial mass of \SI{e8.50}{\solmass} and the seeding distribution $p_2$ was used.
	}
	\label{fig:formationhistory}
\end{figure}

By extracting the redshift values at which GCs are seeded, we are able to make predictions on their age distribution across all simulated haloes (Fig.~\ref{fig:formationhistory}). More than half of all GCs are formed more than \SI{12}{\giga\year} ago, with a peak formation rate above \SI{4}{\per\giga\year\per\mega\parsec\cubed} at \SI{12.9}{\giga\year} ($z=\num{6.4}$). At later times, the formation rate drops quickly. The first 10~per cent of GCs are formed by $z=\num{8.8}$, the first third by $z=\num{5.9}$, and the last third after $z=\num{3.0}$. Only 10~per cent of all GCs are younger than \SI{8.6}{\giga\year} ($z=\num{1.2}$). For the higher seeding mass $\Mseed = \SI{e8.67}{\solmass}$, the total number of GCs in the simulation and the formation rate decrease by a third. However, the redshift values until which the specified fractions of GCs formed increase only slightly by $\Delta z \sim \num{.1}\text{--}\num{.2}$.

\begin{figure}
	\centering
	\includegraphics{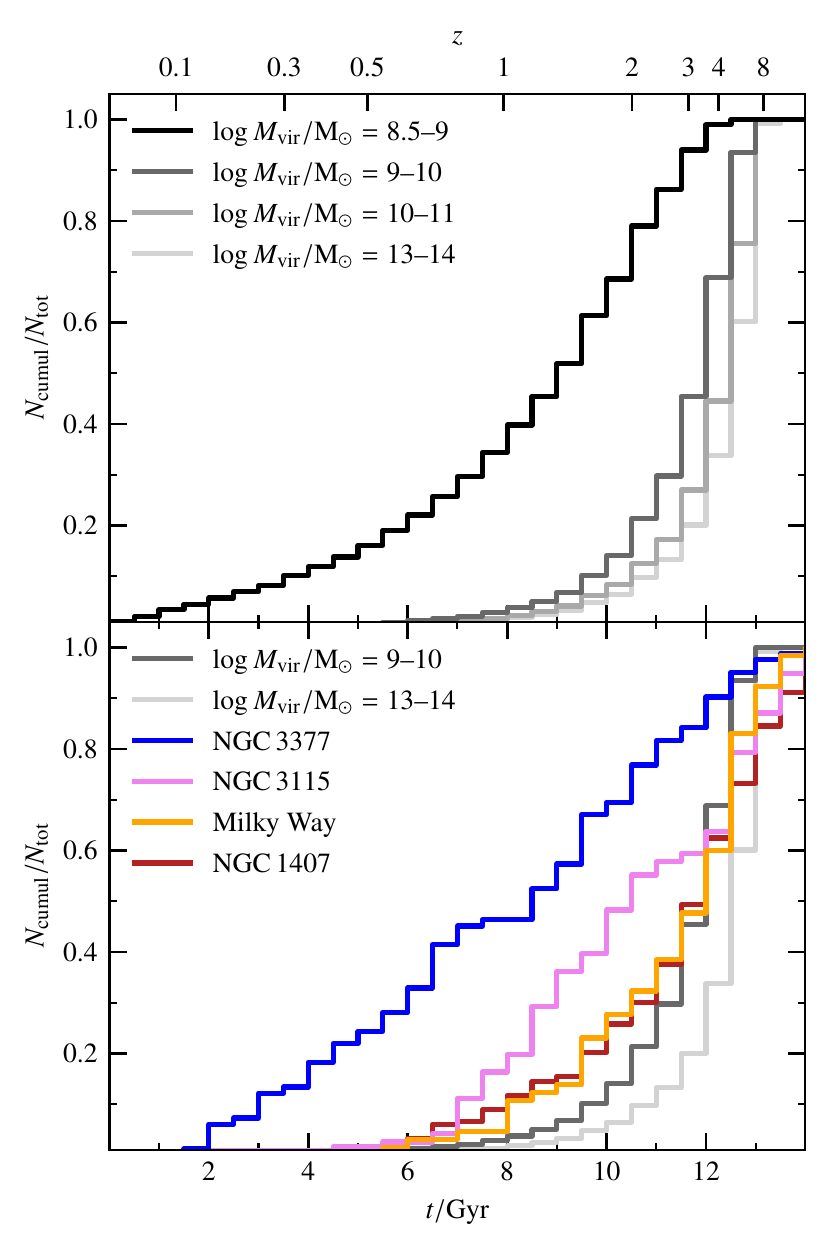}
	\caption{
	    \textit{Top}: Ages of GCs in simulated haloes as a normalised cumulative histogram for different halo virial masses. Haloes with masses between $\log \Mvir/{\solmass} = 11$ and~13 contain GCs with age distributions located between the two lines corresponding to masses right above and below. We find that in general, the larger a halo's mass is, the fewer young GCs it contains. GCs were seeded at a virial mass of \SI{e8.50}{\solmass} and the seeding distribution $p_2$ was used. \textit{Bottom}: Comparison of the four globular cluster age populations provided by \citet{usher:2019} to the simulated populations. The lowest mass bin is not shown because of being both too low for any of the four galaxy haloes and too close to \Mseed{}, thus being strongly affected by the model's formation mechanism.
	}
	\label{fig:agedistribution}
\end{figure}

Binning the dark matter haloes in virial mass, we determined the average GC age distributions for galaxies in those bins (see upper panel of Fig.~\ref{fig:agedistribution}). In general, we find that most of the GCs are old, in good agreement with observations \citep[e.g.][]{forbes:2001,strader:2005,brodie:2006}. More specifically, we find that more massive haloes generally contain older GCs, while less massive haloes contain relatively more younger GCs. Between 30~and 40~per cent of GCs in haloes more massive than \SI{e11}{\solmass} are older than \SI{13}{\giga\year}, with formation times between redshifts $z=4$ and~7. But also for the smaller halo masses between \num{e9}~and \SI{e10}{\solmass}, about half of the GCs are older than \SI{12}{\giga\year}.

The strongest difference can be found between the mass bins $\log M/\solmass = 8.5\text{--}9$ and $\log M/\solmass = 9\text{--}10$: While there already is a lack of GCs younger than \SI{6}{\giga\year} in haloes of the more massive bin, about 20~per cent of the GCs in haloes of the lowest mass bin are younger than \SI{6}{\giga\year}, and about 50~per cent are actually younger than \SI{9}{\giga\year}. On the one hand, this is in agreement with what is seen for the small Magellanic Cloud, where the oldest GCs are only \SI{8}{\giga\year} old, and the youngest have ages below \SI{5}{\giga\year} \citep{parisi:2014}. \citet{sharina:2006} even report two GCs in local dwarf galaxies with ages below \SI{2}{\giga\year}. On the other hand, these haloes have masses barely above our seeding mass of $\Mseed = \SI{3.2e8}{\solmass}$, and are thus strongly affected by the formation criterion implemented in our model. Furthermore, it is also not clear from observations whether GCs are still born in low-mass haloes at low redshifts in the same way as at high redshifts. The reason for the question is that there are also observations of dwarf galaxies with purely old GC systems \citep{strader:2005}, and several with mixed ages \citep{sharina:2006}. However, given that all those dwarf galaxies only have very few GCs, the picture is not conclusive yet.

We compare our results with the GC age distributions presented by \citet{usher:2019} for four different galaxies, NGC\,3377, NGC\,3115, NCG\,1407, and the MW, with the GC ages of the latter taken from \citet{vandenberg:2013} and \citet{leaman:2013}. Their dark halo virial masses are on the order of \num{e12}, \SIlist{e12;e13;e12}{\solmass}, respectively (\citealp{forbes16} for the former three, using their estimate of the total dark matter virial mass being a factor~10 higher than within 8~effective radii; and \citealp{karukes20} for the MW). As can be seen in the lower panel of Fig.~\ref{fig:agedistribution}, the age distributions of the MW and NGC\,1407 are similar to what is found in our model for galaxies with similar halo masses, in particular with respect to the older GCs. However, both observed galaxies have a younger GC population with ages of \SIrange[range-phrase=--]{8}{12}{\giga\year}, which is much less present in our model.
The same age deviation for middle-aged GCs can also be seen for NGC\,3115, but to a larger degree. NGC\,3377 shows the strongest deviation from our results, with an age distribution similar to what we find for the smallest mass bin, though it is much more massive than the haloes in the smallest mass bin. There is not a single simulated halo in the same mass bin as NGC\,3377 with a similar GC age distribution.
We specifically looked at the haloes with the youngest average GC systems with halo masses above \SI{e10}{\solmass} and could not find a single GC age distribution similar to NGC\,3377. While we could find haloes with GC systems that have a young age tail, the GCs with ages less than \SI{6}{\giga\year} were always below 15~per cent. We also looked at the formation histories of the haloes with the youngest and oldest GC systems and could not find a trend with either the amount of mergers nor with the amount of smooth accretion.

We conclude that the simple assumption of GC assembly through hierarchical merging with GCs being born in low mass haloes of about $\Mseed = \SI{3.2e8}{\solmass}$ can well reproduce the general observed trend for galaxies of larger halo masses to also have older GC systems as well as the peaks at old ages above \SI{12}{\giga\year}, as seen for NGC\,1407 \citep{usher:2019} and M87 \citep{cohen:1998}. This clearly indicates that hierarchical merging and mixing is sufficient to explain the overall population of old GCs in galaxies.
We can also find rather young GC populations in the dwarf halo range, albeit this is certainly closely linked to the seeding criterion.
However, we also clearly see that our simple model is not able to account for the observed amount of medium-aged GCs with ages of \SIrange[range-phrase=--]{8}{12}{\giga\year} \citep{peng:2004,usher:2019}, and especially fails to explain the younger GC populations observed in some non-dwarf galaxies like NGC\,3377 \citep{usher:2019} or NGC\,1316 \citep{goodfrooij:2001,sesto:2018}. 
To account for these, an additional GC formation mechanism is necessary, for example forming GCs in gas-rich (major) mergers, which is discussed in the following section.

\section{Dual Formation Pathways}
\label{sec:dualformation}

\subsection{Merger Model Properties}
\label{sec:extension_propertiesj}

\begin{figure}
	\centering
	\includegraphics{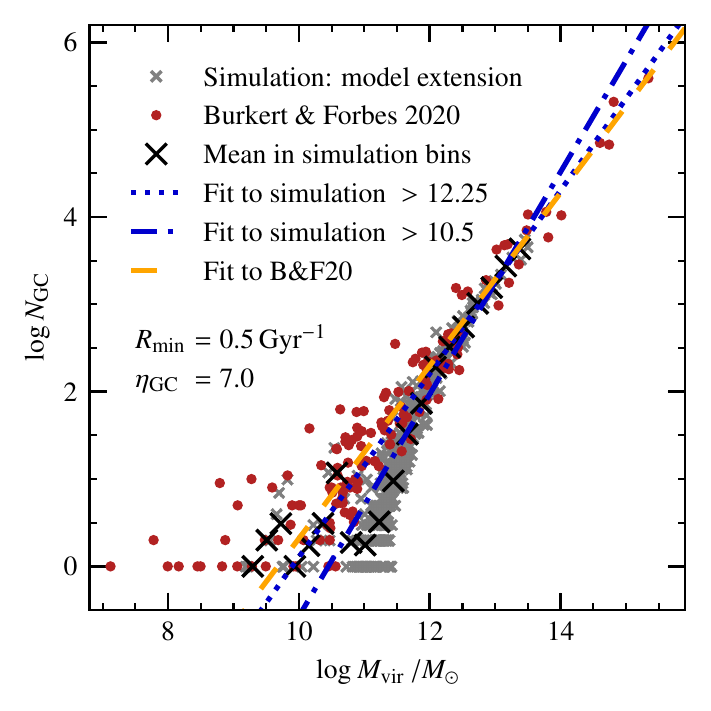}
	\caption{
		Correlation between dark matter halo virial masses \Mvir{} and numbers of globular clusters \NGC{} created through the merger formation pathway compared to the observational sample from \citetalias{burkert20}. The model parameters are $\Amin = \SI{.5}{\per\giga\year}$ and $\etaGC = \num{7.0}$. The values of the simulated virial masses are the true values and are not shifted by mock observation errors. The fits to the simulation were computed by the least-square method for the bin masses above \SIlist{e12.25}{\solmass} and \SIlist{e10.5}{\solmass}, averaged over 1000~mock observations of the simulated virial masses. The fit to the observations corresponds to Equation~\ref{eq:NGCMvirlogburkert}.
	}
	\label{fig:Mvir_NGC_red_extension}
\end{figure}

Because of the clear lack of younger GCs compared to observations (Section~\ref{sec:ages}) and the suggestions in literature of GC formation in gas-rich mergers \citep[e.g.][]{ashman92}, we extended the base GC model by an additional formation mechanism generally triggered by gas-rich mergers (model description in Section~\ref{sec:model_extension}). Before analysing the full combined model, we first inspected the properties of the merger model only, i.e., of the GCs formed through the merger formation pathway alone. Using $\Amin = \SI{.5}{\per\giga\year}$ and $\etaGC = \num{7.0}$, which lie close to the best fit of \citet{choksi:2018} ($\Amin = \SI{.5}{\per\giga\year}$ and $\etaGC = \num{6.75}$), leads to a good resemblance with the observational abundances of \citetalias{burkert20} above $\log\Mvir \sim \num{12.25}$ (Fig.~\ref{fig:Mvir_NGC_red_extension}), in agreement with \citet{choksi:2018} and \citet{boylan17}. However, it is evident that the simulated number of GCs is too low in less massive haloes and GCs are almost non-existent below ${\sim}\SI{e10.5}{\solmass}$ compared to what is observed \citep{forbes18}. This is true for all parameter combinations of \Amin{} and \etaGC{}.

An important feature of the merger model is that the GCs formed in mergers are generally younger than those formed in the base model, which precisely addresses the weak point of the base model in the comparisons with observations. This is what makes the combination of the base model and merger model appealing: GCs are formed both in low-mass haloes and in mergers, and the age distribution is made younger by the GCs formed in the merger model.

\subsection{Parameter Space \& GC Numbers}
\label{sec:parameter_space_extension}

The merger model introduces two further free parameters, \Amin{} and \etaGC{}. Together with \Mseed{}, the parameters are too degenerate with respect to the GC numbers to be able to fit all three of them simultaneously to the observed abundances. Unfortunately, the scatter of \NGC{} does not help in constraining these since the scatter varies too little between parameter sets and is always smaller than the observed scatter, likely due to observational errors of \NGC{} and \Mvir{}. Because of the small number of full observed GC age distributions, these can also only be used as an indicator, but not for any precise fitting procedure. For these reasons, we explored the parameter space by fixing \Amin{} and \etaGC{} to then adapt \Mseed{} to best match the observations of \citetalias{burkert20} and \citetalias{harris17} (Tables~\ref{tab:parameter_space_forbes} and~\ref{tab:parameter_space_harris}). Lower values of the logarithmic halo mass accretion rate threshold, \Amin{}, and higher values of the gas mass conversion parameter, \etaGC{}, lead to more GCs being formed through mergers. The models in which these parameters alone lead to too many GCs being formed compared to the respective observational sample are excluded and denoted by dashes in the tables. Models at the border of having too many of these GCs do not allow for enough GCs to be formed through the base model, thus underpopulating the GCs in low-mass haloes and having an S-shaped \NGC{}-\Mvir{}-relation instead of a linear one. Those models are denoted by values put in parentheses. All other models are good options when only considering the abundances. For some of these models there are gas-rich mergers with up to ${\sim}\num{100}$~GCs being formed at once.

\begin{table}
    \centering
    \caption{
        Fitted base model seeding masses, $\log\Mseed/\si{\solmass}$, to the observations of \citetalias{burkert20} depending on \Amin{} and \etaGC{}. Cells with a dash indicate that the respective models had too many GCs formed through the merger model, even without considering the base model. Values in parentheses indicate that those models had too few GCs in low-mass haloes compared to the observations. Cells highlighted in grey correspond to the best-fitting GC fractions formed in mergers as discussed in Section~\ref{sec:gc_fractions}.
    }
    \label{tab:parameter_space_forbes}
    \begin{tabular}{S[table-format=1.1] *{8}{S[input-symbols=(),table-format=1.2]}}
        \toprule
         & \multicolumn{8}{c}{$\Amin / \si{\per\giga\year}$} \\ \addlinespace
        \etaGC{} & {0.1} & {0.2} & {0.3} & {0.4} & {0.5} & {0.6} & {0.7} & {0.8} \\
        \midrule
        7.0 & {--} & {--} & {--} & (9.20) & (9.00) & (8.90) & (8.90) & (8.80) \\
        6.0 & {--} & {--} & (9.20) & (9.10) & (9.00) & 8.90 & 8.90 & \cellcolor{hc} 8.80 \\
        5.0 & {--} & {--} & (9.10) & (9.00) & 8.80 & \cellcolor{hc} 8.80 & \cellcolor{hc} 8.80 & \cellcolor{hc} 8.80 \\
        4.0 & {--} & (9.00) & (9.00) & 8.90 & \cellcolor{hc} 8.80 & 8.70 & 8.70 & 8.65 \\
        3.0 & {--} & (8.80) & \cellcolor{hc} 8.80 & \cellcolor{hc} 8.75 & 8.75 & 8.70 & 8.65 & 8.65 \\
        2.0 & (8.80) & (8.75) & 8.70 & 8.65 & 8.65 & 8.60 & 8.60 & 8.60 \\
        1.0 & 8.65 & 8.60 & 8.60 & 8.55 & 8.55 & 8.55 & 8.55 & 8.55 \\
        \bottomrule
    \end{tabular}
\end{table}

\begin{table}
    \centering
    \caption{
        Values for $\log\Mseed/\si{\solmass}$ as in Table~\ref{tab:parameter_space_forbes}, fit to the observations of \citetalias{harris17}.
    }
    \label{tab:parameter_space_harris}
    \begin{tabular}{S[table-format=1.1] *{8}{S[input-symbols=(),table-format=1.2]}}
        \toprule
         & \multicolumn{8}{c}{$\Amin / \si{\per\giga\year}$} \\ \addlinespace
        \etaGC{} & {0.1} & {0.2} & {0.3} & {0.4} & {0.5} & {0.6} & {0.7} & {0.8} \\
        \midrule
        5.0 & {--} & {--} & {--} & (9.20) & (9.20) & (9.00) & (9.00) & (9.00) \\
        4.0 & {--} & {--} & (9.20) & (9.10) & (9.10) & (9.00) & (9.00) & 9.00 \\
        3.0 & {--} & (9.30) & (9.10) & 9.00 & 9.00 & \cellcolor{hc} 8.95 & \cellcolor{hc} 8.90 & 8.85 \\
        2.0 & (9.20) & (9.10) & 9.00 & \cellcolor{hc} 8.95 & \cellcolor{hc} 8.95 & 8.90 & 8.85 & 8.85 \\
        1.0 & (9.10) & \cellcolor{hc} 9.00 & 8.95 & 8.90 & 8.80 & 8.75 & 8.70 & 8.70 \\
        \bottomrule
    \end{tabular}
\end{table}

\begin{figure*}
	\centering
	\includegraphics{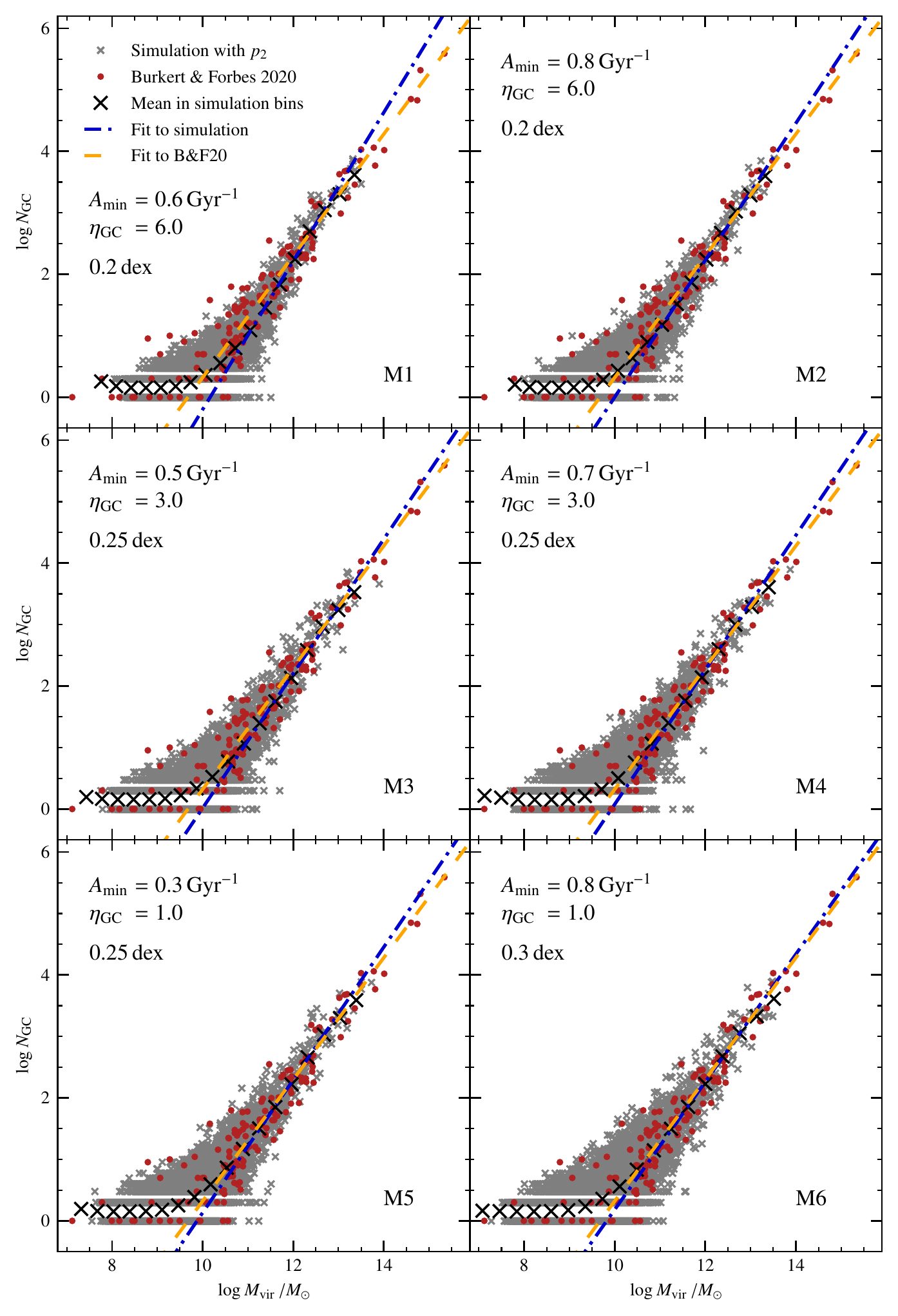}
	\caption{
		Correlation between dark matter halo virial masses \Mvir{} and numbers of globular clusters \NGC{} created through a selection of models (composed of the base and merger models) compared to the observational sample from \citetalias{burkert20}. GCs were seeded at the virial masses indicated in Table~\ref{tab:parameter_space_forbes}. In all cases, the base model seeding distribution $p_2$ was used. The virial masses of the simulations are mock observations of the originally simulated ones, as indicated by the values given in \si{\dex}. The fits to the simulations were computed by the least-square method for the bin masses between \SIlist{e10.5;e13}{\solmass}, averaged over 1000~mock observations of the simulated virial masses. The fits to the observations correspond to Equation~\ref{eq:NGCMvirlogburkert}.
	}
	\label{fig:Mvir_NGC_red_forbes_panel}
\end{figure*}

Simultaneously, we determined the mock observation errors of each model that lead to the best match of the scatter of \NGC{} with the observations.
The GC abundances of a selection of six models fit to \citetalias{burkert20} are plotted in Fig.~\ref{fig:Mvir_NGC_red_forbes_panel} with the values of \Amin{} and \etaGC{} denoted in the respective plots, where the respective mock observation errors are applied to the halo virial masses. \Mseed{} decreases from models M1 to M6 towards the value determined for the base model in Section~\ref{sec:mockoutput}. In the same way, the fraction of GCs formed in mergers decreases approximately equally-spaced from 55 down to 10~per cent in haloes with $\Mvir > \SI{e12.5}{\solmass}$.
We found the trend that having more GCs formed in mergers leads to a steeper linear fit that deviates more from the observational fit. Given that we fit the data starting at \SI{e10.5}{\solmass}, this shows how the slope of the linear fit approaches that of the dash-dotted line in Fig.~\ref{fig:Mvir_NGC_red_extension} for the merger model alone when the number of GCs formed through merger events is increased.

\begin{figure}
	\centering
	\includegraphics{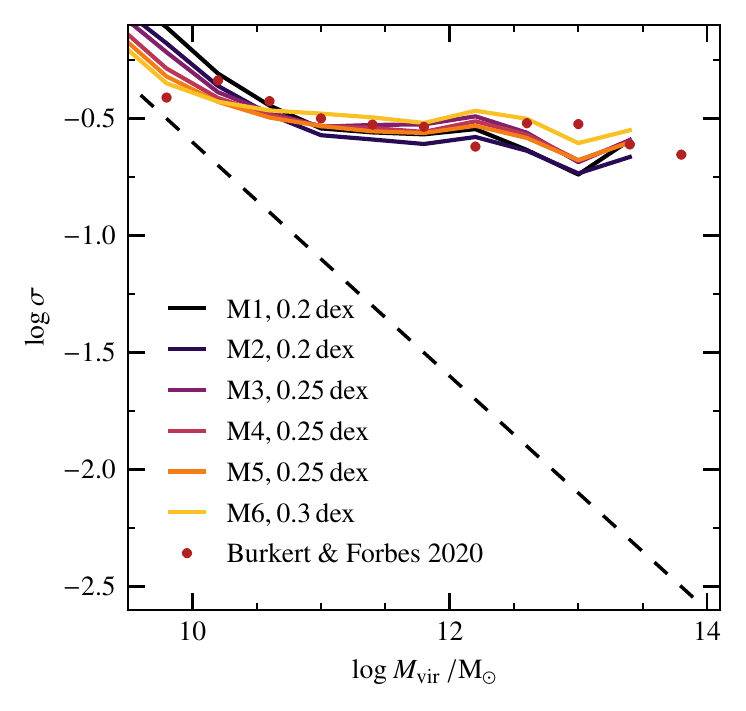}
	\caption{
		Scatter of \NGC{} around the mean correlation compared between a selection of models (same as in Fig.~\ref{fig:Mvir_NGC_red_forbes_panel}) and the observed haloes from \citetalias{burkert20}. The dashed line corresponds to the prediction of the central limit theorem. The solid lines show the average scatter for 1000~mock observations of the respective simulated virial masses according to the values given in \si{\dex}.
	}
	\label{fig:Mvir_sigma_dex_red}
\end{figure}

It is also interesting to note that having more GCs formed in mergers leads to a larger intrinsic scatter of \NGC{} in the models. This can be seen in the required mock observation errors that have to be applied to the simulated virial masses to reproduce the observed scatter (Fig.~\ref{fig:Mvir_sigma_dex_red}). While it needs to be \SI{.3}{\dex} for the model with the least GCs formed in mergers (M6), it drops down to \SI{.2}{\dex} for the models with the largest fraction of them (M1, M2). The same also applies to the models fitted to \citetalias{harris17}. In no case does the scatter come close to the observed one, however.

This shows that further properties that introduce a range of differences in GC formation naturally lead to a larger intrinsic scatter of \NGC{}. This means that our findings simply indicate a likely upper limit of the actual observational uncertainties of \SIrange{.2}{.3}{\dex} (ignoring any systematic errors).

\subsection{Red GC Fractions}
\label{sec:gc_fractions}

\begin{figure}
	\centering
	\includegraphics{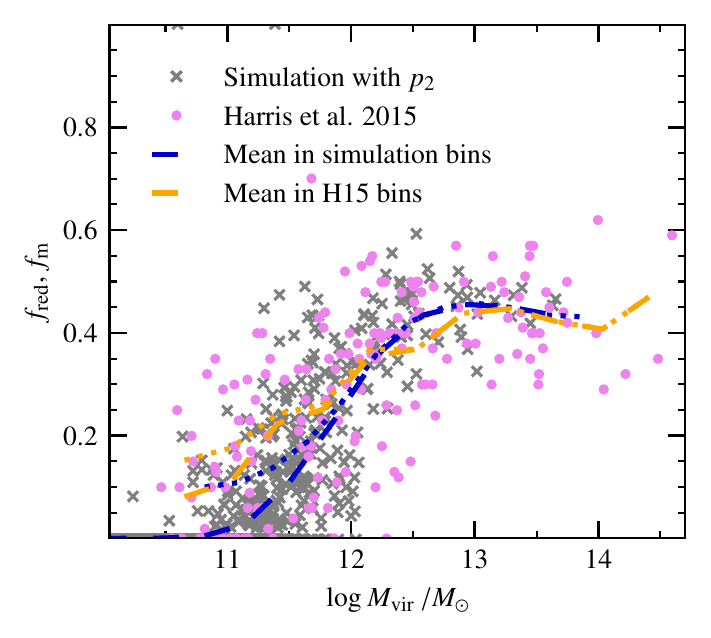}
	\caption{
	    Fraction of GCs formed through the merger model compared to the fraction of red GCs from \citet{harris15}. The parameters used in the simulation were $\Amin = \SI{4.0}{\per\giga\year}$, $\etaGC = \num{.5}$, and $\Mseed = \SI{e8.80}{\solmass}$ with the base model seeding distribution $p_2$. The values of the simulated virial masses are mock observations of the originally simulated values by \SI{.25}{\dex}. The dashed lines indicate the respective means per mass bin, including haloes with fractions of zero. The dotted lines only consider the haloes with non-zero fractions. For the simulation, the bin means were determined by the average over \num{1000}~mock observations of the true simulated virial masses.
	}
	\label{fig:Mvir_frac_red}
\end{figure}

We compared the fractions of GCs formed in mergers with the observed red GC fractions from \citet{harris15}. Under the assumption that GCs formed in the merger model tend to correspond to observed red GCs and that those formed in the base model tend to correspond to blue GCs, we decided to use these fractions to further constrain the parameter space to a rough sensible region. For this, we used the observed feature that the mean fraction of red GCs as a function of \Mvir{} appears to converge at around 40--45~per cent, which is also consistent with recent observations on a small sample by \citet{kang21}. All simulated models also have such a feature, converging towards higher values for larger \Amin{} and smaller \etaGC{}. The models that lead to mean fractions of GCs formed in mergers between 39 and 46~per cent in haloes with $\Mvir > \SI{e12.5}{\solmass}$ are highlighted in grey in Tables~\ref{tab:parameter_space_forbes} and~\ref{tab:parameter_space_harris}. A comparison of one of those models with \citet{harris15} is shown in Fig.~\ref{fig:Mvir_frac_red}, generally demonstrating a good agreement between observational and simulated values. The simulated fractions are lower than those observed for virial masses below ${\sim}\SI{e12}{\solmass}$. Excluding the haloes with fractions of zero results in higher mean values for both the observations and simulations at virial masses below ${\sim}\SI{e11.5}{\solmass}$, but it still does not resolve the difference in the lower mass range. This is also the case for the other models highlighted in the tables. Clearly, the current merger model is not entirely capable of representing the red GC population, especially at low halo masses, although the similar behaviours of the fractions indicate that this approach has its merit.
Overall, the best-fitting models feature base model seeding masses of around $\Mseed = \SI{e8.80}{\solmass}$ for \citetalias{burkert20} and $\Mseed = \SI{e8.95}{\solmass}$ for \citetalias{harris17}, which are about twice the values found for the base model alone.

\subsection{Age Distributions}
\label{sec:ages_extension}

\begin{figure}
	\centering
	\includegraphics{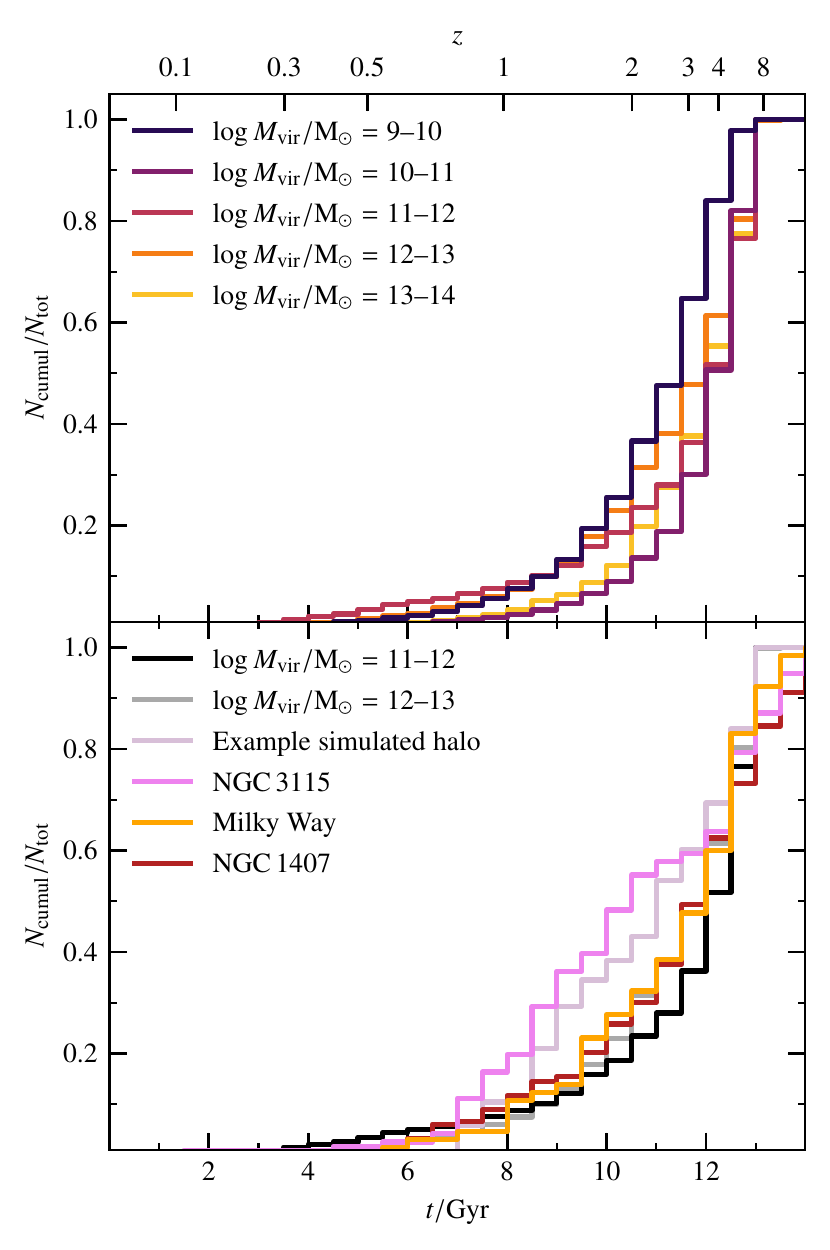}
	\caption{
	    \textit{Top}: Ages of GCs in simulated haloes as a normalised cumulative histogram for different halo virial masses. The parameters used in the simulation were $\Amin = \SI{4.0}{\per\giga\year}$, $\etaGC = \num{.5}$, and $\Mseed = \SI{e8.80}{\solmass}$ with the base model seeding distribution $p_2$. For virial masses until ${\sim}\SI{e11}{\solmass}$, we find that the larger a halo's mass is, the fewer young GCs it contains. This trend is inverted until ${\sim}\SI{e13}{\solmass}$ and then is again valid for even higher virial masses. \textit{Bottom}: Comparison of three of the globular cluster age populations provided by \citet{usher:2019} to the simulated populations in the relevant virial mass bins. The example simulated halo has a virial mass of \SI{e12.3}{\solmass}, contains \num{525}~GCs, and has one of the GC populations that best match that of NGC\,3115.
	}
	\label{fig:agedistribution_extension}
\end{figure}

As for the base model, we extracted the redshift values at which GCs are formed in the different models to obtain the GC age distributions. Gas-rich mergers take place in haloes orders of magnitude more massive than \Mseed{}, therefore leading to an on average later formation time of GCs. Because of this, GCs in massive haloes tend to be younger for larger mean fractions of GCs formed in mergers at the high mass end. Among the selected models with the appropriate fractions of GCs formed in mergers compared to observed red GC fractions (Section~\ref{sec:gc_fractions}), those with lower values of \Amin{} and \etaGC{} have younger GC systems on average. This means that forming many GCs at a time (high \etaGC{}) with a higher accretion threshold (low \Amin{}) leads to GCs being formed at earlier times on average. Comparing the GC age distributions in the respective virial mass bins with the observations from \citet{usher:2019}, we find that the six highlighted models from Tables~\ref{tab:parameter_space_forbes} and~\ref{tab:parameter_space_harris} with $\etaGC \leq \num{.5}$ feature the best match with NGC\,1407 and the MW in the range $\log\Mvir/\si{\solmass} =$ \num{12}--\num{13} (see Fig.~\ref{fig:agedistribution_extension} for one example model).

An important difference to the age distributions found for the base model (Fig.~\ref{fig:agedistribution}) is that now there is an age inversion found in the virial mass range $\log\Mvir/\si{\solmass} =$ \num{11}--\num{13}: instead of the GCs being older with increasing virial mass, the opposite is true in that range (also clearly visible in Fig.~\ref{fig:MvirMeanAge_extension}). The reason for this is the appearance of GCs formed in mergers at around $\Mvir = \SI{e11}{\solmass}$, which generally have younger ages and therefore drop the mean GC age. Interestingly, this inversion is overridden at virial masses above ${\sim}\SI{e13}{\solmass}$, likely due to the decreased rate of gas-rich major mergers in those haloes at late times \citep{oleary:2021}.

In contrast to the base model, the GC age distributions now vary a lot more among haloes with virial masses above \SI{e11}{\solmass}. In fact, we were able to find haloes in some of the models that feature GC populations with similar age distributions to that of NGC\,3115 (bottom panel of Fig.~\ref{fig:agedistribution_extension}), which we could not solely with the base model. Particularly in the models fit to \citetalias{harris17}, where the GC ages are on average younger because of the larger base model seeding mass, there are some haloes that have age distributions very close to the observed one of NGC\,3115. A closer look at the assembly history of the simulated halo best matching the GC age distribution of NGC\,3115 revealed that it has many minor mergers around $z \sim \num{1}$--\num{2} and a rather slow growth before and after. Observers have determined that a likely formation scenario for NGC\,3115 is composed of two phases with an early gas-rich history and a secular evolution thereafter \citep{arnold11,guerou16,poci19,buzzo21}. It is interesting that the selected simulated halo has a history with mergers that appear to take place slightly later than inferred from observations, but still follows the overall picture. This suggests that GC models may be used to extract additional information on the assembly histories of galaxies once their origin is better understood.

However, even in these extended models, we could still not find one halo with GCs similar in age to NGC\,3377, proving that even though this model is now capable of producing GC populations with realistic ages on average, even leading to distributions with younger GCs like NGC\,3115, it still does not cover more extreme cases.

\subsection{Smooth Accretion \& Mean GC System Ages}
\label{sec:smoothaccretion_extension}

\begin{figure}
	\centering
	\includegraphics{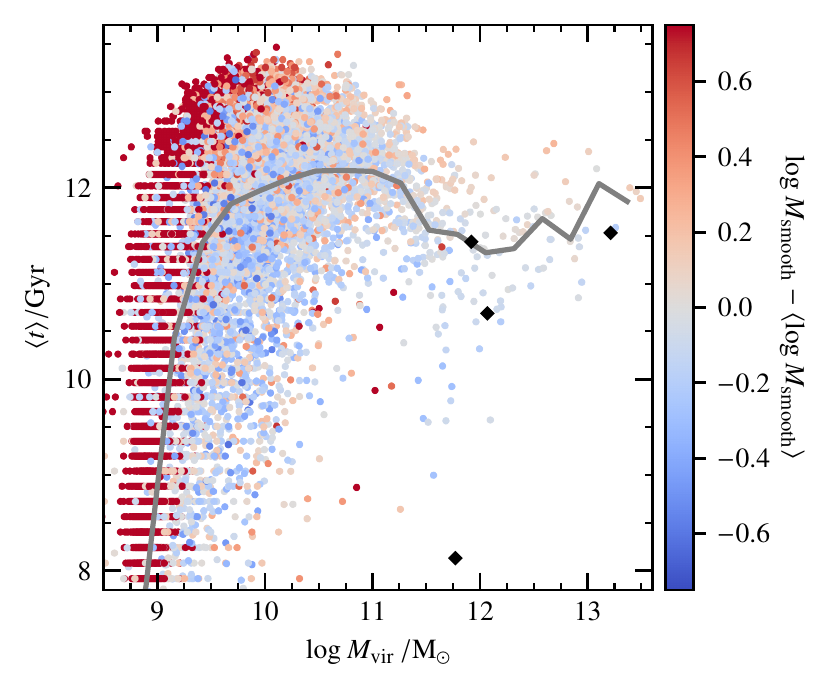}
	\caption{
	    Mean GC ages in the simulated haloes, coloured by the deviation of the total smoothly accreted dark matter mass from the expected value at the respective virial mass. The parameters used in the simulation were $\Amin = \SI{4.0}{\per\giga\year}$, $\etaGC = \num{.5}$, and $\Mseed = \SI{e8.80}{\solmass}$ with the base model seeding distribution $p_2$. The solid grey line indicates the mean over haloes in the virial mass bins. The black diamonds represent the observed galaxies from \citet{usher:2019}, with NGC\,1407 at the top right, and the MW, NGC\,3115, and NGC\,3377, from top to bottom.
	}
	\label{fig:MvirMeanAge_extension}
\end{figure}

Introducing the merger model does not destroy the correlation between smooth accretion and the scatter of \NGC{} discussed in Section~\ref{sec:smoothaccretion}. Additionally, it introduces a further trend visible for the more massive haloes: the older the GC population, the more dark matter was smoothly accreted onto the respective halo compared to similarly massive haloes (Fig.~\ref{fig:MvirMeanAge_extension}). An explanation for this is that the added GC merger model traces haloes with gas-rich major mergers, which also tend to have a smaller amount of smooth accretion, as the model forms more young GCs through the merger events. In contrast, any haloes with more smooth accretion and fewer mergers will have a larger relative amount of GCs formed in low-mass haloes, generally corresponding to the older GC population. Combining the two trends of smooth accretion with \NGC{} and the mean GC age, this means that there is also a trend for haloes with on average younger GCs having slightly more GCs in total.

Fig.~\ref{fig:MvirMeanAge_extension} also shows how the large downward scatter of mean GC system ages at around $\Mvir \gtrsim \SI{e11.5}{\solmass}$ leads to the age inversion discussed in the previous section. The indicated observed galaxies from \citet{usher:2019} lie in this interesting mass range and all but NGC\,3377 have mean GC system ages consistent with the simulated values. The downward scatter is more pronounced the more GCs are formed in mergers and therefore depends on the parameters used for the GC merger model. Further estimations of mean GC system ages in observed galaxies could prove helpful in further analysing the properties of such a downward scatter feature at $\Mvir = \SI{e11.5}{\solmass}$ to better understand GC formation processes in mergers.

\subsection{Dark Matter Halo Mass per GC}
\label{sec:mdmgc_extension}

\begin{figure}
	\centering
	\includegraphics{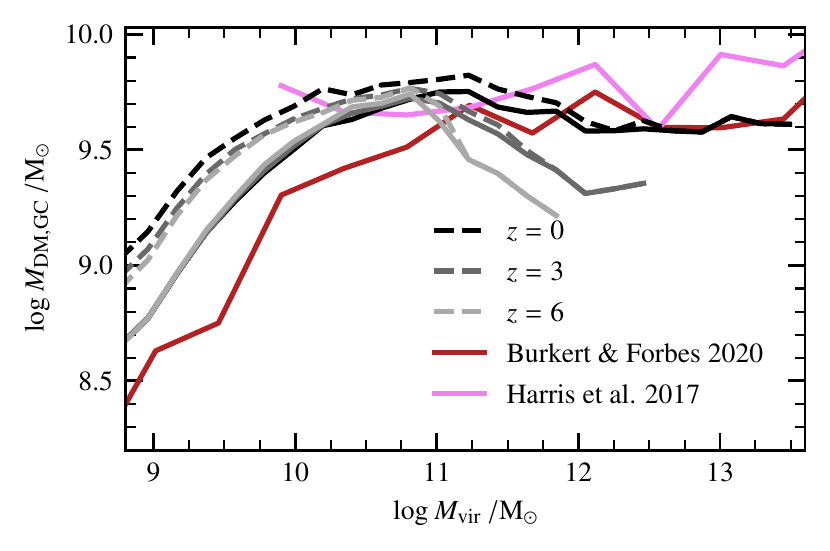}
	\caption{
		Characteristic dark matter mass in virialised haloes per GC for the extended GC model at redshifts $z = 0$, 3, and~6, compared to the samples from \citetalias{burkert20} and \citetalias{harris17}. The dashed lines include haloes without GCs, whereas the solid lines only consist of haloes with at least one GC. The parameters used in the simulation were $\Amin = \SI{4.0}{\per\giga\year}$, $\etaGC = \num{.5}$, and $\Mseed = \SI{e8.80}{\solmass}$ with the base model seeding distribution $p_2$. 
	}
	\label{fig:mdmgc_extension}
\end{figure}

Because of the two different pathways that trigger GC formation at different times and in haloes of different virial masses, the dark matter halo mass per GC for the extended model changes compared to the base model (Fig.~\ref{fig:mdmgc_extension}). Due to the typically later formation of GCs formed through mergers, the number of GCs in the lower virial mass range is smaller than previously, leading to a peak of \MDMGC{} at around \SI{e11}{\solmass}. The peak is less pronounced when only taking haloes with at least one GC into account. Because of this, \MDMGC{} is consistently too high compared to the dwarf galaxy regime from \citetalias{burkert20}. These values decrease towards what is found in the base model the smaller the fraction of GCs formed in mergers is. Another difference is that now the evolution of \NGC{} with redshift changes for high-mass haloes, decreasing with $z$, such that the peak of \MDMGC{} at \SI{e11}{\solmass} is intensified at higher redshift. This means that haloes in the upper virial mass range tend to have slightly more GCs on average at higher redshifts. The general redshift evolution at low halo virial masses remains the same as for the base model (Section~\ref{sec:dwarfgalaxyregime}). The deviation from observations in the dwarf galaxy regime shows that larger observational samples in this halo mass range will be helpful to further constrain the dual formation models. As before, the models fit to \citetalias{harris17} simply lead to higher values of \MDMGC{}, but continue to show the same qualitative features as for the models fit to \citetalias{burkert20}.

\section{Discussion}
\label{sec:discussion}

Our presented base model links the number of globular clusters in a dark matter halo to the halo's virial mass. The resulting distribution of GCs and the linear relationship with halo virial masses is in good agreement with recent observations by \citetalias{harris17} and \citet{forbes18}.
We used the data from \citetalias{burkert20} and \citetalias{harris17} to obtain the seeding mass, i.e., the halo virial mass at which GCs are seeded, and obtained $\Mseed = \num{3.2e8}$ and \SI{4.7e8}{\solmass}, respectively.
We show that the linearity between GC numbers and halo virial masses results from hierarchical merging, and that the distribution of GCs at virial masses above ${\sim}\SI{e11}{\solmass}$ does not depend on the details of GC formation, in agreement with models by \citet{elbadry19} and \citetalias{burkert20}.
Varying the seeding distribution, which determines how many GCs are seeded in a single dark matter halo when it reaches \Mseed{}, therefore proved to have no effect on the resulting relation.
Differences only arise for lower virial masses, in particular in the dwarf galaxy regime, for which observational data are still too limited to draw definite conclusions.
Our choice to use the distribution that seeds \numlist[list-final-separator={, or }]{0;1;2} GCs with equal probabilities therefore only has little impact on our findings. We expect that further observations in the dwarf galaxy regime will give us a better understanding of when GC populations are formed and in what quantities.

Because of generally too old GC populations compared to observed GC age distributions, we extended this base model by a second GC formation pathway following the approach taken by \citet{choksi:2018}. In this \emph{merger model}, GCs are formed when a dark matter halo's virial mass increases at a higher rate than a accretion rate threshold, \Amin{}, according to an assumed cluster initial mass power law function and a gas-to-GC-mass conversion factor, \etaGC{}. This results in GCs generally being formed in gas-rich major mergers for the merger model. These GCs are generally found in haloes with virial masses above \SI{e10.5}{\solmass}. Because of a lack of better observational constraints for the properties analysed in this work, we limited ourselves to a parameter space analysis of the combined models. Constraining the models with observations from \citet{harris15} by assuming a correlation between observed red GCs and the GCs formed through mergers in the simulation, we suggest that values of $\Mseed \sim \num{6.3e8}$ and \SI{8.9e8}{\solmass} are the most likely when fitting \NGC{} to \citetalias{burkert20} and \citetalias{harris17}, respectively, approximately twice as large as for the base model alone. Using the observed GC age distributions from \citet{usher:2019}, we constrained \Amin{} to values below \SI{.5}{\per\giga\year} and $\etaGC \leq \num{4.0}$ and~\num{2.0}, with the upper limit of \etaGC{} depending on the respective observational sample fitted to. While the value of \Amin{} matches the best fit found by \citet{choksi:2018} of $\SI{0.5}{\per\giga\year}$, they find a higher conversion factor of $\etaGC = \num{6.75}$. This deviation is a result of our model additionally consisting of the base model, therefore requiring fewer GCs formed through the merger pathway, while their model tries to capture all observed GCs solely through the merger formation pathway.

There has recently been discussion about the relationship between GCs and the virial masses of their host haloes in the dwarf galaxy regime. \citet{forbes18} found that the relation between the total GC mass in a halo and the halo's virial mass extends linearly down to $\Mvir \sim \SI{e8}{\solmass}$, almost two orders of magnitude lower than observations had previously allowed us to investigate. They noted that they found this proportionality for galaxies with at least one GC.
\citet{boylan17} presented a model to reproduce the same linear relationship, but predicted that the minimum halo virial mass hosting GCs at $z = 0$ is ${\sim}\SI{e10}{\solmass}$, which is not in agreement with recent data. The semi-analytical model introduced by \citet{elbadry19} leads to very little GC formation in haloes less massive than $\Mvir \sim \SI{e10.5}{\solmass}$. They did note, however, that increasing the formation efficiency of GCs in low mass haloes would extend the linear relationship to lower masses. All of these models do not investigate `mean' properties of all haloes, irrespective of having GCs. Instead, they focus on the properties of GC `populated' haloes.
In contrast, our model is capable of both obtaining GCs in the dwarf galaxy regime and also making predictions for the `mean' properties of GC populations in dark matter haloes. However, it fails to agree with the prediction made by \citet{forbes18}, that the mean total GC mass per halo virial mass would decrease below $\Mvir \sim \SI{e10}{\solmass}$ if galaxies without GCs were to be included in the observational data. Instead, our model indicates that the opposite is true: dark matter haloes with low virial masses contain more GCs per halo mass. It should be noted that one difference is that we only count the number of GCs, whereas they used the total GC mass in a halo. Still, we do not see any way in which our model would agree with their prediction, even if we were to model the total mass of GCs instead of using GC numbers.

We find that the scatter of the simulated number of GCs in haloes lies between the prediction of the central limit theorem and the actual observed scatter for all seeding distributions. Varying the seeding distribution does not noticeably increase the scatter of the GC numbers for high virial masses, however. This means that not only the linear relation, but even the scatter of \NGC{} at higher virial masses follows from hierarchical merging, independent of the GC formation details. The merger model adds its own intrinsic scatter in \NGC{}, however only slightly increasing the total scatter when combining the base and the merger model.
Adding a mock observation error to the simulated halo virial masses results in a scatter of the GC numbers that matches the observed scatter more closely. This agrees with the findings of \citetalias{burkert20} that the larger scatter in the observations can be explained by an error of dark matter halo virial masses of a factor of~2.
Adding more properties and features with their own intrinsic scatter to \NGC{} would naturally lower this value, meaning that our findings simply indicate a likely upper limit of the actual observational errors (ignoring any systematic errors).

For the first time, we show that the inherent scatter of the number of globular clusters is closely related to the amount of \emph{smooth accretion}, i.e., a halo's mass that was not accreted through mergers with haloes more massive than \Mseed{}. This relation shows that for $\Mvir \gtrsim \SI{e10}{\solmass}$, haloes with above-average smooth accretion generally contain a below-average number of GCs. Since this inherent scatter is much smaller than that introduced by the mock observations, it would mean that the number of GCs is actually a very precise indicator for the virial mass of the corresponding dark matter halo, particularly for those in the high-mass regime. This is still the case when adding the second GC formation pathway, which even introduces a further correlation with the mean GC age in high-mass haloes: the younger the GC population is, the less dark matter mass was smoothly accreted compared to haloes at the same virial mass, and therefore the more GCs the halo contains. However, these relations could change when allowing for GC disruption or when applying a different mechanism for the second GC formation pathway.

\Mseed{} is an order of magnitude lower than the characteristic dark matter mass in virialised haloes per GC at $z = 0$ found when fitting the linear relation to the observations. This means that the amount of dark matter mass per GC increases with higher masses, reaching a factor of almost ten. This is caused by the smoothly accreted dark matter, which clearly shows that it is important and necessary to account for smooth accretion when modelling GC formation and the relationship between GCs and dark matter halo virial masses. Even for the extended models, the base model seeding masses are still more than half an order of magnitude below the linear relation found at $z = 0$. Additionally, they introduce a peak of the characteristic dark matter mass per GC at $\Mvir \sim \SI{e11}{\solmass}$. This feature will allow further constraining the models with the help of more GC observations in galaxies with haloes in this mass range, as the peak becomes more pronounced the more GCs are formed through mergers.

For the base model, we used the best-fitting values for the GC seeding halo virial mass and the mock observation error to predict what the GC number distribution looks like at higher redshifts. We find that there is no significant difference between redshifts, except that we find less massive galaxies at higher redshifts, attributing to the fact that more massive galaxies need more time to grow through merging and thus only appear at lower redshifts.
The similarities of the relation between \NGC{} and \Mvir{} and its scatter at different redshifts suggest that the merging histories of haloes with equal virial masses, but at different times, is self-similar, which agrees well with the findings of \citet{wechsler02}.
Adding the GC merger model, GCs are formed at different times and in different environments, depending on their formation pathway. This affects the relation between \Mvir{} and \NGC{} at higher redshifts, leading to a slightly larger number of GCs in haloes with virial masses above \SI{e11}{\solmass}.
Note that such observations of GCs at high redshifts are very difficult and may not be possible.

Finally, we also looked at the age distributions of the simulated GCs in the base model, both across all simulated haloes and also for different host halo virial mass ranges. Generally, most of our GCs are old, in agreement with observations \citep[e.g.][]{forbes:2001,strader:2005,brodie:2006}. We find a clear formation peak at $12\text{--}\SI{13}{\giga\year}$, with two thirds of GCs already formed by redshift $z=3$.
We could also successfully reproduce the tendency for more massive haloes to host older GC systems, indicating that this is caused by the hierarchical assembly and not by internal galaxy formation processes.
In addition, there is also a tail of middle aged and young GCs in the simulated sample, albeit the GCs with ages below \SI{6}{\giga\year} are extremely rare.
Even when looking at individual haloes, we could not find any with a significant middle-aged or even young GC population. This is in contrast to observations, which report middle-aged GCs for almost all GC populations for which age distributions are measured \citep[e.g.][]{usher:2019}.

Introducing the second GC formation pathway through the merger model indeed led to generally younger GC populations, simultaneously increasing the variance of GC age distributions. Some of the models in the constrained parameter space feature mean age distributions in the appropriate halo virial mass bins very similar to those found in two of the galaxies analysed by \citet{usher:2019}, the MW and NGC\,1407. Even galaxies with somewhat outlying age distributions, such as NGC\,3115, are matched by GC populations in selected individual haloes from the simulations.
The suggested two-phase formation scenario of NGC\,3115 composed of an early gas-rich merger history and a secular evolution thereafter \citep{arnold11,guerou16,poci19,buzzo21} is overall consistent with that of the matching simulated halo. We suggest that understanding the origin of GCs better will allow using GC models to extract additional information on the assembly histories of galaxies in the future.
However, even the extended model is still incapable of explaining young GC systems, as for example reported for the small Magellanic Cloud \citep{parisi:2014}, NGC\,3377 \citep{usher:2019}, and NGC\,1316 \citep{goodfrooij:2001,sesto:2018}.

An interesting feature that emerges in the extended model is an age inversion in the virial mass range $\log\Mvir/\si{\solmass} =$ \num{11}--\num{13}, where more massive haloes contain overall younger GC populations. Obtaining more observational data on mean GC ages of galaxies' GC populations will help place further constraints on the details of GC formation and their pathways by searching for such an age inversion feature and inspecting how pronounced it might be.

Since this is an empirical model, we have chosen to base it on few but simple assumptions. It currently restricts us to solely making statements about the number of GCs in dark matter haloes, their formation times, and, when including the extension to the base model, the ratios between GCs of different formation pathways. For this work, we decided against trying to model other GC properties, such as masses, metallicities, and any sorts of dynamics, including positions and velocities within the host halo. We also believe it is of importance to more accurately model the relation between GC numbers and virial masses in lower mass galaxies before extending the model with additional properties. It may be worthwhile to investigate possible effects of GC disruption as well in the future.
We conclude that our base model, which assumes GCs to be born in low-mass haloes, is sufficient to explain the origin of the old GCs and a low number of younger GCs, but that an additional channel of GC formation is needed to account for the observed amount of middle-aged and young GCs, as shown by the GC merger model through the formation of GCs during gas-rich merger events (see also \citealp{ashman92} for more details).

\section{Conclusion}
\label{sec:conclusion}

In this work, we presented an empirical model for the number of globular clusters in galaxies by using observational data on the relationship between dark matter halo virial masses and GC numbers to fit the model.
We iterate halo merger trees that are extracted from cosmological dark matter simulations and seed GCs at a certain halo virial mass, \Mseed{}, according to a prespecified GC number seeding distribution. GCs are only formed through this process and cannot disappear once seeded. Haloes that become subhaloes turn over their GC population to the respective main halo.

We also presented an extended model that for the first time combines two different GC formation pathways. The additional formation mechanism depends on a dark matter mass accretion rate threshold, \Amin{}, and typically occurs in gas-rich mergers. GCs are formed according to a conversion prescription from gas to GC mass and a cluster initial mass power law function, meaning that multiple GCs can be formed simultaneously.

We fitted \Mseed{} to achieve the best agreement between our model and the observational data compiled by \citet{burkert20} and \citet{harris17}. The best fit is obtained for $\Mseed = \num{3.2e8}$ and \SI{4.7e8}{\solmass}, respectively. We found that we can reproduce the observed linear correlation between halo virial masses and GC numbers very well, in particular for high-mass haloes. The observed scatter of GC numbers around the linear relationship can be explained by an observation error of a factor of~2 for dark matter halo virial masses. For the extended model, constraining the parameter space leads to base model seeding masses that are approximately twice as large.

The model has allowed us to come to the following key conclusions:
\begin{itemize}
    \item The GC number seeding distribution has no impact on the linear relation between \Mvir{} and \NGC{} or its scatter for $\Mvir \gtrsim \SI{e11}{\solmass}$ because of hierarchical merging.
    \item The correlation and scatter between \Mvir{} and \NGC{} does not change with redshift, except for massive haloes only appearing at lower redshifts. Introducing the second formation pathway introduces a peak of the mean dark matter mass per GC in haloes at around \SI{e11}{\solmass} and decreases at higher redshifts in more massive haloes.
    \item Smooth accretion leads to a characteristic dark matter mass per GC that is one order of magnitude larger than \Mseed{} (half an order of magnitude for the extended model), meaning that GCs form in less massive haloes than expected from the linear relation between \Mvir{} and \NGC{} observed for higher-mass haloes.
    \item The average amount of smooth accretion does not change with redshift at fixed halo virial mass, clearly showing that the different appearance of galaxies at different redshifts is purely driven by gas and star formation physics and not by dark matter accretion.
    \item The inherent scatter of \NGC{} at fixed halo virial mass is strongly correlated with the deviation of smoothly accreted mass from the mean at the given virial mass. However, this scatter is smaller than that introduced by halo mass measurements, making \NGC{} a good tracer for the halo virial mass.
    \item Including the second GC formation pathway introduces a further correlation between the mean GC age and the deviation of smoothly accreted mass, and therefore also between the mean GC age and the scatter of \NGC{} at fixed halo virial mass.
    \item The base model achieves to form old GC populations, but under-populates the young ones. The extended model achieves much better compatibility with observed GC ages, however still fails to reproduce very young GC populations found in some particular galaxies.
    \item The extended model with two GC formation pathways is capable of producing numbers of GCs formed through mergers that follow the observed relation between \Mvir{} and red GC fractions in galaxies very well.
\end{itemize}

We conclude that our presented base model is already sufficient to explain several observed properties of GC systems in galaxies, clearly highlighting the importance of hierarchical merging for the number assembly of GCs. We can also successfully explain the age peak of GCs at high redshifts due to the formation of GCs in low-mass haloes. However, to more accurately reproduce observations with respect to GC ages, the additional second GC formation pathway is necessary. None of the formation mechanisms on their own covers the broad range of properties observationally found for the numbers and ages of GCs in galaxies of different masses.
We believe that our presented model lays the groundwork for even more sophisticated empirical models that are also capable of fine-tuning the distribution of GC numbers in low-mass galaxies and further improving the comparison to red GC populations to best match observations.

\section*{Acknowledgements}

We thank Duncan Forbes for helpful discussions.
We also thank the anonymous referee for their helpful suggestions.
This research was supported by the Excellence Cluster ORIGINS, funded by the Deutsche Forschungsgemeinschaft (DFG, German Research Foundation) under Germany's Excellence Strategy -- EXC-2094-390783311.
BPM and JAO acknowledge an Emmy Noether grant funded by the Deutsche Forschungsgemeinschaft (DFG, German Research Foundation) -- MO 2979/1-1.
Finally, we thank the developers of \code{Astropy} \citep{astropy13,astropy18}, \code{Jupyter} \citep{jupyter16}, \code{Matplotlib} \citep{matplotlib07}, \code{NumPy} \citep{numpy20}, \code{SciPy} \citep{scipy20}, and \code{Pandas} \citep{pandas10,pandas1.1.3} for their very useful free software.
This research has made use of NASA's Astrophysics Data System Bibliographic Services and of the arXiv preprint repository.

\section*{Data Availability}

The data underlying this article will be shared on reasonable request to the corresponding author.




\bibliographystyle{mnras}
\bibliography{bib}



%
%


\bsp	
\label{lastpage}
\end{document}